\documentclass[apjl]{emulateapj}

\usepackage{lscape}

\shorttitle{The Type~Ib supernova 2007Y}
\shortauthors{Stritzinger et al.}
\usepackage{graphicx}
\usepackage{epstopdf}

\received{2009 Jan.  19}
\accepted{2009 Feb. 3}
\begin{document}
\title{The He-rich core-collapse supernova 2007Y: Observations from 
X-ray to Radio Wavelengths\altaffilmark{1,2}}

\author{
Maximilian Stritzinger, \altaffilmark{3,4}
Paolo Mazzali,\altaffilmark{5,6}
Mark~M.~Phillips,\altaffilmark{3}
Stefan Immler,\altaffilmark{7,8}
Alicia Soderberg,\altaffilmark{9,10}
Jesper Sollerman,\altaffilmark{4,11}
Luis Boldt,\altaffilmark{3}
Jonathan Braithwaite,\altaffilmark{12}
Peter Brown,\altaffilmark{13}
Christopher R.~Burns,\altaffilmark{14}
Carlos Contreras,\altaffilmark{3}
Ricardo Covarrubias,\altaffilmark{3}
Gast\'on Folatelli,\altaffilmark{15}
Wendy~L.~Freedman,\altaffilmark{14}
Sergio Gonz\'alez,\altaffilmark{3}
Mario Hamuy,\altaffilmark{15}
Wojtek Krzeminski,\altaffilmark{3}
Barry~F.~Madore,\altaffilmark{14,16}
Peter Milne,\altaffilmark{17}
Nidia Morrell,\altaffilmark{3}
S.~E.~Persson,\altaffilmark{14}
Miguel Roth,\altaffilmark{3}
Mathew Smith,\altaffilmark{18}
and
Nicholas~B.~Suntzeff\altaffilmark{19}
}

\altaffiltext{1}{
 This paper includes data gathered with the 6.5 meter Magellan telescope 
at Las Campanas Observatory, Chile.}

\altaffiltext{2}{Partly based on observations collected at the European Southern
Observatory, La Silla and Paranal Observatories, Chile (ESO Programme 078.D-0048 and
380.D-0272.)}

\altaffiltext{3}{Las Campanas Observatory, Carnegie Observatories,
  Casilla 601, La Serena, Chile; {mstritzinger@lco.cl, mmp@lco.cl, lboldt@lco.cl,
 ccontreras@lco.cl, ricardo@lco.cl,  sgonzalez@lco.cl,
wojtek@lco.cl, nmorrell@lco.cl, miguel@lco.cl}.}

\altaffiltext{4}{Dark Cosmology Centre, Niels Bohr Institute, University 
of Copenhagen, Juliane Maries Vej 30, 2100 Copenhagen \O, Denmark; {max@dark-cosmology.dk, jesper@dark-cosmology.dk}.} 

\altaffiltext{5}{Max-Planck-Institut f\"ur Astrophysik, Karl-Schwarzschild-Str. 1, 85741 Garching bei M\"unchen, Germany; {mazzali@mpa-garching.mpg.de}.}

\altaffiltext{6}{INAF - Osservatorio Astronomico di Padova, vicolo dell'Osservatorio 5, 35122 Padova, Italy.}

\altaffiltext{7}{Astrophysics Science Division, X-RayAstrophysical Labora- 
tory, Code662, NASAGoddardSpaceFlight Center, Greenbelt, MD 20771, USA;
{stefan.m.immler@nasa.gov}.}

\altaffiltext{8}{Universities Space Research Association, 10211 Wincopin Circle, Columbia, MD 21044, USA} 

\altaffiltext{9}{Department of Astrophysical Sciences, Princeton University, Ivy Lane, Princeton, New Jersey 08544, USA; {alicia@astro.princeton.edu}.}

\altaffiltext{10}{Harvard-Smithsonian Center for Astrophysics,
60 Garden Street, Cambridge, MA 02138, USA.}

\altaffiltext{11}{Department of Astronomy, Stockholm University, AlbaNova, 
SE-10691 Stockholm, Sweden.}

\altaffiltext{12}{Canadian Institute for Theoretical Astrophysics, 60
St.\,George St., Toronto M5S 3H8, Canada; {jon@cita.utoronto.ca}.}

\altaffiltext{13}{Pennsylvania State University, Department of Astronomy \& 
Astrophysics, University Park, PA16802, USA; {pbrown@astro.psu.edu}.}    

\altaffiltext{14}{Observatories of the Carnegie Institution of
 Washington, 813 Santa Barbara St., Pasadena, CA 91101; 
 {cburns@ociw.edu, wendy@ociw.edu, barry@ociw.edu, persson@ociw.edu}.}

 \altaffiltext{15}{Universidad de Chile, Departamento de Astronom\'{\i}a,
  Casilla 36-D, Santiago, Chile; {gaston@das.uchile.cl, mhamuy@das.uchile.cl}.}

\altaffiltext{16}{Infrared Processing and Analysis Center, Caltech/Jet
  Propulsion Laboratory, Pasadena, CA 91125.}  

\altaffiltext{17}{Department of Astronomy and Steward Observatory, University of Arizona, Tucson, AZ 85721, USA; {pmilne@as.asrizona.edu}.}

 \altaffiltext{18}{Cosmology and Gravity Group, Department of Mathematics and Applied Mathematics, University of Cape Town, South Africa; {mathew.smith@uct.ac.za}.}
 
   \altaffiltext{19}{Texas A\&M University, Physics Department, College
Station, TX 77843; {nsuntzeff@tamu.edu}.}

\begin{abstract}
  \noindent  
  
   A detailed study spanning approximately a year has been conducted on the Type~Ib supernova 2007Y.
Imaging was obtained from  X-ray to radio wavelengths, and a comprehensive set of multi-band ($w2m2w1u'g'r'i'UBVYJHK_s$) light curves and optical spectroscopy is presented. A virtually complete bolometric light curve is derived, from which we infer a $^{56}$Ni-mass  of 0.06 M$_{\sun}$.
The early spectrum strongly resembles SN~2005bf and exhibits high-velocity features of \ion{Ca}{2} and H$\alpha$; during late epochs the spectrum shows evidence of a 
ejecta-wind interaction.
Nebular emission lines have  similar widths and exhibit profiles that indicate
a lack of major asymmetry in the ejecta.
Late phase spectra are modeled with  a non-LTE code, from which we find $^{56}$Ni, O  and total-ejecta masses (excluding He) to be 0.06, 0.2 and 0.42 M$_{\sun}$, respectively, below 4,500 km~s$^{-1}$.
The $^{56}$Ni mass confirms results obtained from the bolometric light curve. The oxygen abundance suggests the progenitor was most likely
a  $\approx$3.3 M$_{\sun}$ He core star that evolved from a
zero-age-main-sequence mass of 10--13 M$_{\sun}$.
The explosion energy is determined to be $\approx$10$^{50}$ erg, and the mass-loss
rate of the progenitor is  constrained from X-ray and radio observations to be $\lesssim10^{-6}$~M$_{\sun}$~yr$^{-1}$.
SN~2007Y is among the least energetic normal Type~Ib supernovae ever studied.
\end{abstract}


\section{INTRODUCTION}
\label{sec:intro}
\noindent 
 A decade has passed since it was first recognized that some
extragalactic gamma-ray transients 
could be linked to the death of massive stars  \citep{galama98}.
Soon afterwards the connection between long gamma-ray bursts (GRBs) and energetic Type~Ic supernovae (SNe~Ic) was firmly established  \citep{hjorth03,matheson03,stanek03,malesani04}. 
Over the last several years it has also become clear that some X-ray flashes, a less energetic version of GRBs,  are produced by a similar type of SN~Ic  \citep{fynbo04,modjaz06,pian06,sollerman06}.

Recently,  X-ray emission was detected from the  He-rich
Type~Ib SN~2008D  \citep{soderberg08,mazzali08,malesani09,modjaz08}.
This finding has given fresh impetus to the study of normal SNe~Ib, which has been somewhat neglected, only a handful of events having been well-observed.  
In this article we report detailed observations, obtained through the course of the 
Carnegie Supernova Project (here after CSP; Hamuy et al. 2006), of the 
Type~Ib SN~2007Y. 
The earliest spectra of this event bear a striking resemblance  to the unusual Type~Ib SN~2005bf \citep{anupama05,folatelli06}, however its light curves exhibit a more 
ordinary evolution. 
  
SN~2007Y (see Fig.~1) was discovered in the nearby spiral galaxy NGC~1187
on Feb.~15.77 UT\footnote[20]{Universal Time (UT) is used through out this article.} \citep{monard07}, and was initially classified as a young peculiar SN~Ib/c \citep{folatelli07}.
As this was a  young and nearby supernova it made an excellent target for our follow-up program.  

 In this article a comprehensive set of X-ray, ultraviolet, optical, near-infrared  and
 radio observations of SN~2007Y are presented to gain insight into the explosion physics, and determine the stellar evolutionary path of the progenitor.
 X-ray and ultraviolet observations were carried out with the
X-Ray Telescope (XRT) and the UltraViolet Optical Telescope  (UVOT) aboard the 
{\em Swift} satellite \citep{roming05}. Early-phase spectroscopy and imaging was 
collected at Las Campanas Observatory (LCO).
Late-time spectroscopy and imaging came from facilities at LCO
and at the European Organisation for Astronomical Research in the Southern Hemisphere's (ESO) Paranal and La Silla Observatories. Finally, our radio observations
were obtained with the Very Large Array.\footnote[21]{The Very Large Array is a
facility of the National Science Foundation operated under
cooperative agreement by Associated Universities, Inc.}

In $\S$~2  a concise description of the observations is given.
More complete details regarding the data reduction methods applied to the optical
and near-infrared observations can be found in the Appendix. Section~3 contains the analysis of the data, while  the discussion is  presented in $\S$~4. We conclude 
with a summary in  $\S$~5.

\section{Observations}

Six weeks of imaging was obtained at LCO covering the flux evolution
from  $-14$ to $+$41 days past maximum light.\footnote[22]{Maximum light refers 
to the time of peak bolometric brightness (L$_{\rm max}$), i.e. March 3.5 or
JD-2454163.12
(see $\S$~\ref{section:bolometric}).}
Optical imaging was performed with a set of Sloan  $u'g'r'i'$ and 
Johnson $B$ and $V$ filters, while in the near-infrared a set of $JHK_s$ filters 
identical to those used by \citet{persson98} were employed. The $Y$ band is defined
in  \citet{hillenbrand02}, and details concerning its calibration are given in 
\citet[][hereafter Paper~I]{hamuy06} and \citet{contreras09}.  
In addition, two epochs of  optical and near-infrared
imaging were conducted at late epochs with the VLT.

Nineteen epochs of early phase photometry obtained with UVOT and originally published in \citet{brown08} are also included in our analysis. These data were obtained with 
$w2m2w1UBV$ passbands and are critical to calculating a nearly complete 
bolometric  light curve. 
SN~2007Y was also observed with XRT. No X-ray emission was detected, but these observations allow us to place an upper limit on the mass-loss rate of the 
progenitor prior to explosion. 

Early and late phase radio monitoring spanning from $-$8 to 653 days past maximum
were obtained with the VLA. 
With these observations constraints are placed on the mass-loss rate of the progenitor 
star and the interaction of circumstellar material (CSM) with the supernova shock wave.
Our radio observations are complementary  to the XRT observations and provide 
a more stringent limit on the mass-loss rate.

Table~1 contains the positions and average magnitudes of the optical and near-infrared local sequence stars used to compute final photometry from the CSP images. 
The definitive optical and near-infrared photometry in the standard \citet{landolt92} 
($BV$) and \citet{smith02} ($u'g'r'i'$) systems are given in 
Table~2 and Table~3, respectively.
 
Early phase ultraviolet, optical, and near-infrared light curves of SN~2007Y are 
shown in Fig.~\ref{fig:lightcurves}. 
Late phase photometry is combined with the corresponding early epoch light curves in  Fig.~\ref{fig:sn07Y_LTphot}.  Note, the VLT optical images were 
taken with a set of facility Johnson Kron-Cousins $BVRI$  and the near-infrared
images with $J_{s}HK_{s}$ filters \citep{persson98,hillenbrand02}.
  
 A total of twelve epochs of long slit optical spectroscopy were collected over the course 
 of  $\sim$300 days. 
 Table~4 contains the journal of spectroscopic observations and gives details
 concerning the telescope, instrument setup, and final data product. The 
 spectra corrected to the rest frame of the host galaxy are presented in Fig.~\ref{fig:spectra}.
For clarity each spectrum has been shifted below the top one by an arbitrary constant. 
The number in parentheses gives the epoch with respect to L$_{\rm max}$.
Four of these spectra were taken when the ejecta was fully nebular. A subset of these
are modeled in order to derive physical  parameters of SN~2007Y 
(see $\S$~\ref{section:modeling}).

\section{Analysis}

\subsection{Distance and reddening}

With coordinates of $\alpha$ = $03^{h}02^{m}35\fs92$ and 
 $\delta$ = $-22\degr53\arcmin50.1\arcsec$,  SN~2007Y was located
  24$\arcsec$ west and 
 123$\arcsec$ south of the nucleus of NGC~1187; well outside of any spiral arm.
 The NASA/IPAC Extragalactic Database (NED) lists a heliocentric recessional 
velocity for this galaxy of 1390~km~s$^{-1}$. 
With a value of H$_{\circ} =$ 72$\pm$8 km~s$^{-1}$~Mpc$^{-1}$ 
\citep{freedman01} and adopting a 400~km~s$^{-1}$ uncertainty in the redshift distance, the 
distance to NGC~1187 is 19.31$\pm$5.56 Mpc ($\mu$ = 31.43 $\pm$ 0.55).
If this distance is corrected for Virgo infall it becomes 10\% smaller. In this paper
we will use the more traditional heliocentric distance, but note that
a Virgo infall corrected distance would make our estimated values of the peak  absolute magnitude, luminosity, and mass of  $^{56}$Ni  about 10\% less. 

According to the infrared dust maps of \citet{schlegel98} the reddening due to 
the Milky Way in the direction of NGC~1187 is E($B-V$)$_{\rm gal} =$ 0.022 mag.
The optical colors of SN~2007Y near peak brightness (see below) 
are relatively blue, suggesting minimal host galaxy extinction. 
A spectrum obtained near maximum light shows the presence of a 
weak \ion{Na}{1}~D absorption feature at the
redshift of the host galaxy with an equivalent width of 0.56~\AA.
Using the relationship between the equivalent width of the 
 \ion{Na}{1}~D and E(B-V) \citep[see][]{turatto03,taubenberger06} suggests
 a host galaxy reddening of E($B-V$)$_{\rm host}$ $=$ 0.09 mag. Combining 
 this with the Galactic reddening gives E($B-V$)$_{\rm tot}$ $=$ 0.112 mag. This
 value is adopted in the subsequent analysis.

\subsection{X-Ray Observations and Limiting Luminosity}
Swift X-Ray Telescope (XRT; Burrows et al. 2005) observations were obtained simultaneously with the UVOT observations. To search for X-ray emission from SN~2007Y, we extracted X-ray counts from a circular region with a 10~pixel radius ($23\farcs7$, corresponding to the XRT on-axis 90\% encircled energy radius) centered on the optical position of the SN. The background was extracted locally from a source-free region of radius of $40''$ to account for detector and sky background, and for diffuse emission from the host galaxy. 

No X-ray source is detected in the merged 66.5~ks XRT data obtained in photon-counting mode. The $3\sigma$ upper limit to the XRT net count rate is $2.63 \times 10^{-4}~{\rm cts~s}^{-1}$, corresponding to an unabsorbed (0.2--10~keV band) X-ray flux of $f_{0.2-10} < 1.24 \times 10^{-14}~{\rm erg~cm}^{-2}~{\rm s}^{-1}$ and a luminosity $L_{0.2-10} < 5.5 \times 10^{38}~{\rm erg~s}^{-1}$ for an adopted thermal plasma spectrum with a temperature of $kT = 10~{\rm keV}$ \citep[see][and references therein]{fransson96}, a Galactic foreground column density of $N_{\rm H} = 1.91 \times 10^{20}~{\rm cm}^{-2}$ \citep{dickey90} and a distance of 19.31~Mpc.

In order to search for a possible late X-ray turn-on phase, we binned the XRT data into two time intervals ranging from 
day $-14$ to $+23$
and day $+$290 to $+$387 
with exposure times of 56.5~ks and 10.1~ks, respectively. No X-ray source is found at the position of SN~2007Y in either of the two epochs, down to $3\sigma$ limiting luminosities of $<6.5 \times 10^{38}~{\rm erg~s}^{-1}$ and $<1.9 \times 10^{39}~{\rm erg~s}^{-1}$, respectively. 

The lack of X-ray emission can be used to put loose limits on the mass-loss rate of the progenitor system.
Following the discussion in \citet[][and references therein]{immler07}, an upper limit to the mass-loss rate of $\dot{M} \approx 2\times10^{-5}$~M$_{\sun}~{\rm yr}^{-1}$~
(v$_{\rm w}$/10~km~s$^{-1})$ with an uncertainty of a factor of 2--3 is obtained. 
In this calculation the model assumes thermal emission, which is more appropriate  for Type~II SNe. In addition the wind velocity of Wolf-Rayet stars 
is known  to be of the order of 1000 km~s$^{-1}$. 
Taking this into account then implies a limit on the mass-loss of $\sim2\times10^{-3}$
 M$_{\sun}$ yr$^{-1}$. To obtain a more robust estimate on the mass-loss rate 
 we turn to constraints provided by the radio observations (see $\S$~\ref{radio}).

\subsection{Ultraviolet, Optical and Near-infrared Light Curves}
\label{lightcurves}
The nicely sampled ultraviolet, optical, near-infrared light curves presented in 
Fig.~\ref{fig:lightcurves} constitute one of the more complete collections 
obtained for a He-rich SN~Ib.

Under close inspection the CSP and UVOT $B$- and $V$-band light curves
exhibit excellent agreement with one another. However, the $U$-band light curve of 
UVOT is systematically $\sim$0.7 mag  brighter than CSP's $u'$-band light curve.
This is not surprising considering the appreciable difference between the transmission functions of these two filters. The $u'$-band has an effective wavelength  $\lambda_{eff}$ $=$ 3681 \AA~and a FWHM of 487 \AA, while  the
$U$-band has $\lambda_{eff} = $ 3465 \AA~ and a FWHM of 785 \AA. 

To  see if this discrepancy could be attributed to the differences in  
filter transmissivities we performed an experiment using spectrophotometry of SN~2007Y. 
First, appropriate zero-points were computed for each passband using a
library of spectrophotometric flux standards \citep{stritzinger05}. 
Armed with these zero-points and the early phase spectra that 
had their wavelength coverage extended to the atmospheric cutoff by assuming
zero flux, synthetic magnitudes of SN~2007Y were 
determined for the two passbands. 
We find the $U$-band filter
consistently gives synthetic magnitudes brighter than $u'$-band, and
the disparity is in agreement with the average difference 
between the broad-band magnitudes. 

 Figure~\ref{fig:color} shows the ($u'-B$) and ($B-V)$ color curves of SN~2007Y, 
 both of which track photospheric temperature variations. Included for comparison 
 are the color curves of  two well-observed SNe~IIb: SN~2008ax \citep{pastorello08} and SN~2008aq (CSP paper in preparation). Also included is the (B-V) color curve
 of the Type~Ib SN~1999ex \citep{stritzinger02}. 
 The color curves of each SNe have been corrected for extinction using reddening values of
 E($B-V$) of 0.30, 0.11, 0.05 and 0.30 for SN~1999ex, SN~2007Y, SN~2008aq and SN~2008ax, respectively.
As the brightness of the supernova increases the photospheric temperature 
also increases, which in turn, causes the color curves to evolve to the blue.
Specifically,  the color of SN~2007Y decreases from a value of ($B-V$) $=$ 0.47 on day
 $-$12.8 to ($B-V$) $=$ 0.06 on day $-4.3$. The colors then evolve monotonically back to the red where ($B-V$) $=$ 1.0 on day $+$31.2. Then the colors moved marginally back towards the blue.
Interestingly, around 3 weeks past maximum the UVOT ultraviolet light curves appear to stop declining in brightness. Although UVOT observations were discontinued near this epoch the $u'$-band light curve indicates that the flux in this passband not only stopped declining but actually increased(!).  Specifically, the $u'$ band  
decreases from 19.800 mag on day $+$22 to 19.408$\pm$0.110 
nineteen days later.
This phenomenon coincides with a ($u'-B$) color evolution to the blue that
is more prevalent than in other He-rich core-collapse SN like SN~2008ax 
(see bottom panel of Fig.~\ref{fig:color}).  
A blue excess is also seen in SN~2008aq where the  $u'$-band light curve becomes
brighter by 0.63 mag from day +$30$ to $+$96. This is followed by a dimming of 
the $u'$ light curve by 0.55 mag from day  $+$96 to $+$123.
Notice in the bottom panel of Fig.~\ref{fig:color} the ($u'-B$) colors curves 
of SN~2007Y and SN~2008aq are very similar between days $+25$  and $+35$.

The excess in ultraviolet emission could be connected to a change in 
opacity of the ejecta. Unfortunately the spectra do not extend far enough 
in the blue to determine if  the spectral energy distribution of the supernova
actually increases in the region covered by the $u'$-band filter.
Alternatively, this could be the consequence of shock heating produced from the interaction of high-velocity supernova ejecta with
CSM \citep[e.g.][]{immler06}. As we will see below there
is evidence in the late epoch optical spectrum of such an interaction.

We now turn to the broad-band absolute magnitudes of SN~2007Y.
First the time and observed magnitude at maximum light of each filter light curve
was estimated with the use of moderate order (5--8) polynomial fits; 
these values are given in Table~5. 
Similarly to other SNe~Ib/c, the ultraviolet/blue  light curves of SN~2007Y peak
$\sim$4 days prior to the optical light curves. This is followed by the near-infrared 
bands peaking 3--5 days later. 
Peak absolute magnitudes were computed using the reddening value of E($B-V$)$_{\rm tot}$ $=$ 0.112 mag, an R$_{V}$ value of 3.1, and a distance of 19.31$\pm$5.56 Mpc
(see $\S$~3.1). These values are also listed in 
Table~5 with the quoted errors accounting for uncertainty in the estimated value of peak magnitude and the distance to the host galaxy.
With an absolute magnitude of M$_{V} =$$-$16.5$\pm$0.6 mag, SN~2007Y is similar in brightness to SN~2008D, which peaked at M$_{V}  \approx$$-$16.7 mag \citep{soderberg08}. 
This turns out to be on the faint end of the absolute magnitude distribution of SNe~Ib/c. 
For example SN~2006aj peaked at  M$_{V} = $$-$18.7$\pm$0.1 \citep{modjaz06}, while other previously observed GRB/X-ray transient related SNe~Ib/c have peaked at 
M$_{V}<-$20.0 mag \citep[see][and references therein]{richardson06}.

In Fig.~\ref{fig:photcompare} the $B$- and $V$-band light curves of 
SN~2007Y are compared with those of four other well-observed SNe~Ib/c: SN~1994I \citep{richmond96}, SN~1998bw \citep{galama98}, SN~2005bf \citep{folatelli06}, and SN~2006aj \citep{mirabal06,sollerman06}.
The observed magnitudes of the  comparison SNe~Ib/c have been normalized to 
the peak magnitudes of SN~2007Y and shifted in time to their respective epoch of peak
brightness.
The light curves of SN~1998bw and SN~2006aj have been corrected for time dilation, and SN~2006aj 
has also had its underlying host galaxy contribution removed. 
SN~2007Y rises to maximum like SN~1998bw and SN~2005bf, however, 
after maximum it declines rapidly like SN~2006aj. 
Then about three weeks past maximum the rate of decline begins to slow,
and again, the $V$-band light curve of SN~2007Y is similar to  
that of SN~1998bw.  We note that SN~1999ex and SN~2008D were not included in  
Fig.~\ref{fig:photcompare}, but if they were, 
one would see that the width of their light curves is 
 significantly broader than SN~2007Y. 
Overall the shape of the $V$-band light curve, which contains the 
majority of flux around maximum, is most similar to SN~2006aj. 

Looking at \citet{arnett82}'s analytical description of the shape of the  light curve around peak brightness, we can infer from the similar shaped light curves  that the effective diffusion time $\tau_{\rm m}$, which can expressed as
\begin{equation}
\tau_{\rm m} \propto \kappa_{\rm opt}^{1/2}M_{\rm ejc}^{3/4}E_{\rm kin}^{-1/4}~~,
\end{equation}
\noindent  is similar in SNe 2006aj and 2007Y.
 In this expression $\kappa_{\rm opt}$ refers to 
the mean optical opacity, M$_{\rm ejc}$ is the ejecta mass, and E$_{\rm kin}$ is 
the kinetic energy of the expansion.
 This then implies the ratio of  M$_{\rm ejc}^3$  to E$_{\rm kin}$  is also similar between
 these two events.

Now we turn our attention to the late phase VLT photometry shown in 
Fig.~\ref{fig:sn07Y_LTphot}.
The optical  light curves indicate magnitude differences from peak to 
the last observed epoch  of $\Delta$m($B) = 7.9$ and 
$\Delta$m($R) = 6.9$. 
Furthermore a weighted least square analysis gives decline rates 
per hundred days of  $B =$ 1.30$\pm$0.31,  $V = 1.73\pm0.20$, $R = 1.65\pm0.25$ 
and $I = 1.82\pm0.25$. Clearly these decline rates are faster than the decline of 0.98
magnitude per hundred days that is  expected in the case
of complete trapping of the $\gamma$ rays produced from the decay of $^{56}$Co to $^{56}$Fe.
Although the measured magnitude decline rates are  rather uncertain given that only two epochs were 
observed, they are nonetheless in agreement with other SNe~Ib/c
\citep[e.g.][]{barbon94,sollerman98,patat01,elmhamdi04,tomita06,clocchiatti08},
and  hints towards a low ejecta mass.
 
\subsection{Bolometric Light Curve}\label{section:bolometric}

The comprehensive wavelength coverage provided by the early-time observations affords
an excellent opportunity to construct a nearly complete bolometric light curve of SN~2007Y.  To this end it was first necessary to account for gaps in any individual 
light curve when a particular passband was not observed on a given night.
In these cases it proved necessary to  interpolate between the existing light curve using low order polynomial fits. 
As the $m2$-band light curve covers only a limited time around maximum its temporal
coverage was extended by extrapolation assuming a color correction (based on the available photometry) of ($w2-m2) = -$0.929.
Similarly as the $K_{s}$ light curve sampling was rather sparse a 
($V-K_{s}$) $=$ 1.0 color was adopted to extend the light curve during the pre-maximum
phase, and a ($V-K_{s}$) $=$ 0.2 color was adopted to extrapolate the light curve from 
day  $+$12 to $+$42. Although these two assumptions are rather simple they do 
provide a reasonably accurate extrapolation of the light curves.

Next, the photometry was corrected for extinction using our adopted value of 
reddening and a R$_{V} =$ 3.1 extinction curve for the optical and near-infrared data, 
while the extinction curve of \citet{pei92} was used to correct the 
ultraviolet photometry.
The magnitudes were then converted to flux at the effective wavelength of each filter.
In the case of the UVOT photometry, zero-points and  count-rate-to-flux
conversion factors given in Table~6  and Table~9 of \citet{poole08} were used. 

To compute the quasi-bolometric (UV to NIR, hereafter UVOIR) light curve the observed magnitudes of each light curve were
converted to flux and then integrated over frequency.
 The summed flux was then converted to luminosity using our adopted distance to NGC~1187. Note the last four epochs of the early phase UVOIR light curve were computed 
 without ultraviolet observations, and the late phase
 bolometric fluxes were derived with $BVRIJ_{s}H$ photometry.
In Fig.~\ref{fig:bolometric} the UVOIR light curve of  SN~2007Y is compared to the UVOIR light curves of several low luminosity SNe~Ib. These include: SN~1999ex \citep{stritzinger02}, SN~2005bf \citep{folatelli06}, SN~2008D \citep{malesani09} and  
the Type~IIb SN~1993J \citep{richmond94}.
Extinction values of A$_{V} =$ 0.6, 0.93 and 2.5 mag, and distances  of 3.6,  44.1 and 30.0 Mpc were adopted for SN~1993J, SN~1999ex and SN~2008D, respectively.
In the case of SN~2005bf,  values of  `L$_{\rm bol}$'  were taken directly from Table~2 of \citet{folatelli06}.

The UVOIR light curve of SN~2007Y indicates a peak luminosity of 
$\sim$1.30$\times$10$^{42}$ erg~s$^{-1}$. 
The main uncertainty in this estimate is the error in the distance which 
is of the order of 30\%. 
One way to estimate 
the abundance of $^{56}$Ni, the main radioactive isotope  that powers 
the light curves of supernovae, is through the use of Arnett's Rule \citep{arnett82}.
Although based on a number of  underlying assumptions,  \citet{richardson06} have used
this method to obtain reasonable estimates of the $^{56}$Ni yields for 
 a sample of  two dozen SNe~Ib/c. In the case of SN~2007Y, if we  
assume a rise time of 18 days, application of Arnett's Rule 
tells us  $\approx$0.06$\pm$0.02 M$_{\sun}$ of $^{56}$Ni was synthesized 
in SN~2007Y.\footnote[23]{We note in passing that the rise times of SNe~Ib vary 
significantly. The rise times of SNe~1999ex and 2008D 
were 18 days \citep{stritzinger02,modjaz08}, while 
SN~2006aj exhibited a rise time of 11 days \citep{sollerman06}.} 

An alternative method of estimating the $^{56}$Ni yield is to fit a radioactive
decay energy deposition function to the late phase UVOIR light curve.
Under the reasonable assumptions that the majority of energy deposited 
in the ejecta at late phases is from the $^{56}$Co $\rightarrow$ $^{56}$Fe decay, 
and that the optical depth, $\tau$, is much less than  one,
a simple model that estimates the $^{56}$Ni-mass can be fitted to the UVOIR light curve from 50 to 350 days. The functional form of the model used 
\citep[see][]{sollerman98} is  $L~=~1.3 \times 10^{43}\, M_{\rm Ni}~\mathrm{e}^{-t/111.3}  \left(1 - 0.966\mathrm{e}^{-\tau}\right)$ $\rm erg~s^{-1}$, where $\tau$ is given by  (t$_{1}$/t)$^{2}$.
In this form  t$_{1}$ sets the time when the optical depth to $\gamma$ rays is unity. 

Before the energy deposition function could be used to estimate the $^{56}$Ni mass
 it first deemed necessary to extend the UVOIR light curve from day $+$41 to 
$+$60. 
 This was achieved by applying a linear least squares fit to the UVOIR light curve from day $+28$ to $+41$; this best fit was then used  to  extrapolate the light curve to day 
 $+$60. Note the assumption that the decline rate between days $+$28 to  $+$41
 can be used to extend all but the $u'$-band light curves out to day $+$60 is reasonable as 
 other SNe~Ib/c  show no changes in decline rate during the epochs
in question  \citep[see][]{clocchiatti08}.
 Finally the best fit of the toy model to the UVOIR light curve was obtained with 
a   t$_{1}$ of 35.8 days and a  $^{56}$Ni mass of 0.06 M$_{\sun}$. 

The broad wavelength coverage obtained for SN~2007Y also allows for an
estimation of the distribution of flux (as a function of time) in the
different wavelength domains.
These estimates are shown in Fig.~\ref{fig:sed_compare}.
Similar to Type~Ia supernovae at maximum the majority of flux is emitted in the optical,  however the ultraviolet light curves contained a significant portion of flux ($\gtrsim$20\%); while the near-infrared contribution was small, $\sim$5\%.
By two weeks past L$_{\rm max}$ the flux in the near-infrared passbands increased to roughly 20\% while in the ultraviolet it dropped to $\leq$10\%.
During the last two epochs (day $+$270 and $+$344) $\sim$40\% of the observed flux
was contained in the $R$ and $I$ bands because these passbands 
coincide with the strong emission lines of \ion{O}{1}, H$\alpha$, \ion{Fe}{2}, and \ion{Ca}{2} that dominate the spectrum (see below). By day $+$270 $\sim$5\% of the flux was contained in the $H$ band while the near-infrared as a whole contributes no more 
than 15\%. Evidently at this phase near-infrared emission lines were not a major coolant.

\subsection{Early Phase Spectroscopic Evolution}

The early spectra of SN~2007Y (Fig. \ref{fig:spectra}) 
consist of a series of P-Cygni features superimposed  on a
blue pseudo continuum.  
Multiple absorption and emission features of \ion{Fe}{2} dominate the blue-end of  the spectra while an exceptionally strong \ion{Ca}{2} $\lambda\lambda$8498, 8542, 8662 triplet rounds out the red-end.
Distinctive features include: (i) an absorption trough between 4000 and
4200~\AA\ caused by the blending of Fe-group emission lines  (i.e. Sc~II, Ti~II or alternatively  Fe~II), (ii) absorption at $\sim$5900 \AA\ caused by a blend of 
the \ion{Na}{1} $\lambda\lambda$5890, 5896 doublet with He~I $\lambda$5876, 
(iii) a distinctive absorption feature at $\sim$6200~\AA\ that is possibly formed from 
a blending of \ion{Si}{2} with high-velocity H$\alpha$, and (iv) a relatively 
weak \ion{O}{1} $\lambda$7774 absorption feature.
  
As SN~2007Y evolved through maximum, the strength of the absorption at 6200 \AA\ 
steadily declined until it disappeared around 10 days past L$_{\max}$.
Meanwhile the depth of the \ion{Ca}{2} triplet also 
decreased until day $+$6 when it transformed back to a  deep absorption.
This phenomenon reflects the transition of the \ion{Ca}{2} triplet being formed from a high-velocity component to a photospheric component (see Fig.~\ref{fig:spectracompare}, and Folatelli et al. 2006 and Parrent et al. 2007). 
After maximum, conspicuous lines of \ion{He}{1} $\lambda\lambda$4471, 4921, 5016, 6678, 7065, 7281 emerge and  begin to dominate the spectrum. 

Figure~\ref{fig:velocities} displays the time evolution of the blue-shifts  
(v$_{\rm exp}$) for the ions that produced the dominant features in the optical spectrum of SN~2007Y. 
The expansion velocity of \ion{Fe}{2} $\lambda$6159 is often used as an indicator of the 
photospheric velocity \citep{branch02,richardson06}. 
In the earliest spectra the absorption minimum of this   
line indicates that v$_{\rm exp}$$\sim$10,000 km~s$^{-1}$, while \ion{He}{1} 
is measured as v$_{\rm exp}$$\sim$7000 km~s$^{-1}$. This 
is reminiscent of SN~2005bf, which to date has been the only 
other SN~Ib to show \ion{Fe}{2} absorption blue-shifted more than its \ion{He}{1}
absorption.    
As SN~2007Y evolved in time, \ion{Fe}{2} $\lambda$6159 monotonically declined, 
reaching a minimum value of $\sim$4500  km~s$^{-1}$ three weeks past maximum.  
On the other hand, the blue-shift of \ion{He}{1} $\lambda$5876 shows an increase of 
$\sim$1500 km~s$^{-1}$ from day $-14$ to L$_{\rm max}$.  
Thereafter the blue-shift of \ion{He}{1} $\lambda$5876 evolved from 8500 to 
6500 km~s$^{-1}$ over the next month. 

Contrarily to all other ions, the velocity at maximum absorption of  
the 6200~\AA\ feature  (assuming it is related to H$\alpha$, 
see $\S$~\ref{discussion}) and the \ion{Ca}{2} triplet, suggest these features 
were produced from a component of gas that was detached from the photosphere 
with v$_{\rm exp}$ $\sim$15,500 km~s$^{-1}$.  
By a week past maximum H$\alpha$ had decreased to  v$_{\rm exp}$ 
$\sim$10,000 km~s$^{-1}$, while the \ion{Ca}{2} triplet began to follow 
the velocity evolution of \ion{He}{1} with v$_{\rm exp}$ $\sim$ 6500  km~s$^{-1}$.
The abnormally low blue-shifts of  \ion{Si}{2} argues against the case that 
it is the dominate ion responsible for the formation of the 6200~\AA\ feature.

In Fig.~\ref{fig:spectracompare} spectra of SN~2007Y at days $-$14, $-$8, $-$1 and $+$39
are compared to spectra of SN~2005bf  obtained on days $-$8, $-$3, $+$4 and 
$+$42.\footnote[24]{Epochs of SN~2005bf spectra are with respect to the time 
of its {\it first} maximum.}
The similarity between the earliest spectra of these two SNe~Ib is remarkable.
Two subtle differences exist, namely a somewhat stronger 
\ion{Ca}{2} triplet in SN~2007Y, while the  \ion{Fe}{2} multiplet 42  
$\lambda\lambda$4924, 5018, 5169 features in SN~2005bf are more narrow. 
In this figure one can see the  position of maximum absorption of the \ion{Ca}{2} triplet clearly  evolves in time.
The narrow features in SN~2005bf, on the other hand, have been interpreted 
to be from a high-velocity  component of Fe gas in the ejecta \citep{anupama05,folatelli06,parrent07}, that maybe related to a polar outflow 
or a failed jet \citep{folatelli06,maeda07b,maund07}. 
The lack of these narrow features in SN~2007Y 
suggests its ejecta was more spherical than in the case of SN~2005bf. 

Finally, at the bottom of  Fig.~\ref{fig:spectracompare}, spectra of SNe~2007Y and 2005bf taken over a  month past maximum are compared. 
By this time the spectra of both events resemble normal SNe~Ib.    

\subsection{Nebular Phase Spectroscopy}

Months later, as SN~2007Y continued to expand, the photosphere receded 
deep into the ejecta leading to the onset of the nebular phase.
 As radioactive heating and the formation of the nebular spectrum
 occurs at this time in what was the central region of the progenitor, the analysis and spectral 
 synthesis of the late phase spectrum provide an excellent tool to estimate the abundances of stable and radioactive elements synthesized during the explosion. Furthermore,
 the study of line profiles and their relative strengths can provide 
clues regarding the geometry of the ejecta, and the nature of the progenitor system.

The four nebular spectra at the bottom of Fig.~\ref{fig:spectra}
exhibit many of the features typically associated with normal SNe~Ib/c. 
Strong emission lines from forbidden transitions of 
 [\ion{O}{1}]  $\lambda$$\lambda$6300, 6363  
and [\ion{Ca}{2}] $\lambda\lambda$7291, 7324 clearly dominate the spectrum.  
The [\ion{Ca}{2}] emission is actually blended with [\ion{Fe}{2}] 
$\lambda$$\lambda$7155, 7172, 7388, 7452 and possibly [\ion{O}{2}] $\lambda$7300.
Other less pronounced features identified include: \ion{Mg}{1}] $\lambda$4570,
the \ion{Na}{1} $\lambda$$\lambda$5890, 5896 doublet, the \ion{Ca}{2} triplet, 
and a feature at 5200~\AA\ that is formed by a  blending of [\ion{Fe}{2}] lines.
A marginal detection of \ion{O}{1} $\lambda$7774 is discernable,
while the identification of \ion{O}{1} $\lambda$5577 is weak to nonexistent.
Gaussian fits to the emission profiles indicate  full-width-half-maximum (FWHM) velocities between 5000 to 7500  km~s$^{-1}$. 
 
 Plotted in Fig.~\ref{fig:lineprofiles} are the emission line profiles,  in velocity space, 
 of  \ion{Mg}{1}], \ion{Na}{1}, [\ion{O}{1}] $\lambda$6300.5,  and 
 [\ion{Ca}{2}] $\lambda$7307.5.
 Clearly these profiles show, in velocity space,  no strong deviations from their 
 rest wavelength, and all but \ion{Na}{1} have a single peaked profile. 
 The \ion{Na}{1} profile, on the other hand, exhibits a moderately flat-topped profile.
  The shape and velocities of these line profiles can provide some insight into
 the distribution of the emitting material \citep{fransson87}, 
 and the geometry of the explosion \citep{maeda08}. 
 As the emission lines are observed to have
similar widths, and that the profiles are symmetric and not sharply
peaked, we conclude that there are no major asymmetries in the ejecta.
 
 The most striking characteristic in the nebular spectra of SN~2007Y is a box-like emission profile red-ward of the [\ion{O}{1}] $\lambda$$\lambda$6300, 6363 doublet. Superposed on this broad component are several narrow emission lines close 
to the expected position of H$\alpha$. 
We identify these emission features as H$\alpha$, [\ion{N}{2}] $\lambda$6584
and  [\ion{S}{2}] $\lambda$6717,
which are associated with an \ion{H}{2} region in the vicinity of SN~2007Y.
We measure a FWHM velocity of the narrow H$\alpha$ component of
$\leq$450 km~s$^{-1}$.
The broad component, on the other hand, exhibits a velocity for the bulk of material to be located between  9000 and 11,000  km~s$^{-1}$.  This material is undoubtedly 
hydrogen gas ejected from SN~2007Y.  
  
 Although rarely seen in other SNe~Ib,  a similar  feature was also present in 
 the Type~Ib SN~1996N \citep{sollerman98}, and more notably in the well-studied transitional Type~IIb supernova (SN~IIb) 1993J \citep{clocchiatti94,filippenko94,patat95,matheson00}.
 A similar feature was also reported in nebular spectra of SN~2005bf
\citep{soderberg05,maeda07b}.
Figure~\ref{fig:LTspectracompare} shows a comparison of
SN~2007Y (day  $+$230) to  SN~1993J (day $+$236) and SN~1996N (day $+$221).
Despite differences in profile shapes and blue-shifts, 
these spectra are quite similar overall. 
During this epoch the H$\alpha$ shoulder seems to be relatively stronger in SN~2007Y.
In the case of SN~1993J the strength of this broad emission feature relative to the
[\ion{O}{1}] doublet continued to evolve in time until 
$\sim$500 days after the explosion. By this epoch the ratio of the strengths of these two features reached unity \citep{matheson00}.

The source feeding the majority of emission in this broad wing is not radioactive heating, but rather, high-velocity hydrogen gas interacting with dense CSM  (see \cite{houck96} and \cite{chevalier94} for details on the physics of this emission process). 
The CSM consists of mass shed by the  progenitor,
while the high-velocity hydrogen gas came from a thin shell of material
 ejected by the supernova itself.  
The velocity of the hydrogen gas measured from the nebular spectrum
is consistent with that derived for H$\alpha$ from the 6200 \AA\ absorption feature
nine days past maximum.   

 Interestingly, as is evident from Fig.~\ref{fig:LTspectracompare}, the strength of
[\ion{Ca}{2}] $\lambda7307.5$ relative to the [\ion{O}{1}] $\lambda$6300 doublet is comparable for all three supernovae.   
Specifically, the [\ion{Ca}{2}]/[\ion{O}{1}]  ratio of the three spectra in  
Fig.~\ref{fig:LTspectracompare}, is 0.6, 0.9, and 1.0 for SN~1993J,  SN~1996N and SN~2007Y, respectively. 
For perspective, this ratio, was $\approx$0.5 in SN~1998bw.
\citet{fransson87,fransson89} have shown that the  [\ion{Ca}{2}]/[\ion{O}{1}] 
ratio stays relatively 
constant during the nebular phase, and is a good diagnostic of the core mass, 
where a higher ratio indicates a less massive core. 
From stellar evolution it is well known that the core mass is highly dependent on the
zero-age-main-sequence mass (ZAMS) of the progenitor. 
That the [\ion{Ca}{2}]/[\ion{O}{1}]
ratio in SN~2007Y is $>$ 0.5 suggests that its
 progenitor star had a main sequence mass comparable to SNe 1993J and 1996N, i.e. M$_{\rm ZAMS}$
 $\lesssim$ 20 M$_{\sun}$ \citep{nomoto93,pods93,shigeyama94}.
 This is much lower than 
 the progenitor of SN~1998bw, which had a main sequence mass 
 M$_{\rm ZAMS}$ $\approx$ 40 M$_{\sun}$ \citep{iwamoto98,nakamura01}. 
 To obtain a more quantitative estimate of the ejecta mass and hence the 
 ZAMS mass of the progenitor we turn to spectral synthesis. 
 
\subsection{Nebular Spectral Synthesis}
\label{section:modeling}

To model the nebular spectrum we used a one-zone non-local thermodynamic equilibrium (NLTE) code \citep{mazzali01,mazzali07a}. 
Heating within the ejecta comes from the deposition of $\gamma$ rays
and positrons produced from the decay of $^{56}$Co to $^{56}$Fe.
This is balanced by cooling via nebular line  emission. The emission rate
of each line is determined by solving a NLTE matrix of level populations
\citep{axelrod80}. The emitting volume is assumed to be of finite extent,  
homologously expanding,  
and has a  uniform density and composition. The emission spectrum is
computed from assigning a parabolic profile to each line which is bounded 
by the velocity of the outer edge of the ejecta.

The main parameters determining the synthetic spectrum are the width of the emission lines, the mass of radioactive $^{56}$Ni, and the mass of the other elements. As iron is a product of the 
decay of $^{56}$Ni, the width of the [\ion{Fe}{2}] emission at 5200 \AA\ gives 
an indication of the distribution of $^{56}$Ni. 
 Abundances of all elements except sulphur and silicon were determined by fitting the emission line profiles. Unfortunately neither of these ions have a strong or isolated 
 emission lines in the optical. Therefore the abundances of these two elements are not
 strongly constrained.

 Given the short baseline in time covered by the nebular spectra we elected to model the two best spectra. These observations and the corresponding synthetic spectra are shown in Fig.~\ref{fig:synthesis}. In both cases the 5300 \AA\ [\ion{Fe}{2}] blend is best fitted with a model having an outer velocity of 4500 km s$^{-1}$. This is also true for the 
[\ion{O}{1}] $\lambda$6300 doublet. 
Having an identical velocity for these two lines is often the case in low energy SNe~Ib/c  \citep[e.g.][]{sauer06}, and is another indication that the explosion was not 
strongly aspherical. 
Both synthetic spectra shown in  Fig.~\ref{fig:synthesis} have a total
 ejected mass of 0.44 M$_{\sun}$ below 4500 km s$^{-1}$. 
 Of this mass 0.06 M$_{\sun}$ is $^{56}$Ni and $\approx$0.20 M$_{\sun}$ is oxygen.
 The value for the $^{56}$Ni abundance confirms the results obtained from 
 both the early and late phase UVOIR light curve.

\subsection{Radio Observations and Circumstellar Interaction}
\label{radio}
In Table~6 we summarize our radio observations of
SN\,2007Y. 
All observations were conducted with the VLA in the
standard continuum mode with a bandwidth of $2\times 50$ MHz centered
at 8.46 GHz.  We used 3C48 and 3C286 (J0137+331 and
J1331+305) for flux calibration, while J0240-231 was used to monitor
the phase.  Data were reduced using standard packages within the
Astronomical Image Processing System (AIPS).

No radio emission was detected at the optical SN position in any of
our VLA observations.  The initial radio upper limits imply that
SN\,2007Y was a factor of $\sim 10^4$ less luminous than SN\,1998bw on
a comparable timescale \citep{kulkarni98}.  We conclude that SN\,2007Y,
like the majority of SNe Ib/c, did not produce relativistic ejecta
along our line-of-sight \citep{soderberg06}.  Our later observations at
$t\sim 2$ yrs further constrain the presence of a relativistic jet
initially directed away from our line of sight.  The upper limit on
the late-time flux density is $F_{\nu} < 54~\mu$Jy ($2\sigma$), a
factor of $10^4$ less luminous than well-studied radio afterglows on
the same timescale, and following the transition to non-relativistic
and spherical Sedov-Taylor expansion (e.g. GRB\,030329;
\citealt{frail05}).

While optical data probe the thermal emission from the slow moving
bulk material, non-thermal synchrotron emission is produced as the
forward shock races ahead and dynamically interacts with the
CSM.  Making the standard assumption that the
post-shock energy density is in equipartition between relativistic
electrons ($\epsilon_e$) and amplified magnetic fields ($\epsilon_B$),
the velocity of the radio emitting material is robustly determined
\citep{chevalier98}.  Adopting typical values, $\epsilon_e=\epsilon_B=0.1$,
provides an estimate for the density of the CSM and, in turn, the 
mass-loss rate of the progenitor star.  Together, the velocity of the
shock (v$_s$) and the mass-loss rate of the progenitor ($\dot{\rm M}$)
determine the luminosity and peak time of the bell-shaped radio
light curve.

Since any radio emission from SN\,2007Y fell below the VLA detection
threshold, we use ($2\sigma$) upper limits from Table~6
to constrain the characteristics (luminosity, shape) of the
light curve.  Adopting the formalism of \citet{chevalier06} for the
parameterization of the radio light curves of SNe dominated by
synchrotron self-absorption, we use our light curve constraints for
SN\,2007Y to place robust constraints on v$_s$ and $\dot{\rm M}$. In
Figure~\ref{fig:vd} we show the two-dimensional v$_s-\dot{M}$
parameter space that is ruled out by our radio upper limits.  With the
necessity that the forward shock velocity is faster than the observed
photospheric velocity in the homologous outflow, a second region is
excluded.  Combining these constraints we find that the mass-loss rate
of the SN\,2007Y progenitor star is less than $\dot{\rm M}\approx
10^{-5}$~M$_{\odot}$~yr$^{-1}$ where we have assumed a
 progenitor wind velocity of $v_w=1000~\rm km~s^{-1}$, appropriate for
 Wolf-Rayet stars.   Adopting a typical shock velocity of
 $v_s\approx 40,000~\rm km~s^{-1}$ consistent with well-studied radio
 SNe Ib/c, implies that $\dot{\rm M}\lesssim 10^{-6}$~M$_{\odot}$~yr$^{-1}$
 which is on the low end of the observed values for radio SNe Ib/c
 \citep{chevalier06}.  Finally we estimate the mass of CSM swept up by the 
forward shock, M$_{\rm CSM}$, over the course of our radio observations  
to be less than M$_{\rm CSM}\approx 2\times 10^{-5}$~M$_{\odot}$ assuming
a standard stellar wind density profile.

\section{Discussion}
\label{discussion}

The observational properties of SN~2007Y have a number of qualities in common with 
other Type~Ib and Type~IIb supernovae. Spectra obtained prior to maximum light 
closely resemble  early spectra of SNe~1999ex  and 2005bf, 
 while at late-times SN~2007Y has  commonalities with SNe~1993J and 1996N.
Nevertheless, subtle differences between these events and SN~2007Y do exist and
reflects, among other things, variations in the composition and mass of the progenitor, the kinetic energy of the explosion, and the degree of asphericity of the ejecta.

The early spectra exhibit the infamous 6200~\AA\ absorption line that
is often, but not always, observed in SNe~Ib/c \citep{matheson01}. 
Currently, it is a matter of open debate which ion(s) form this feature.
 Previous authors have proposed a number of  sources that range
from H$\alpha$ \citep{branch02,anupama05,folatelli06,parrent07,pastorello08}
or  
 \ion{Si}{2} $\lambda$6355 \citep[e.g.][]{hamuy02}, to  \ion{Si}{2} blended with 
 H$\alpha$ \citep{tanaka08}. 
 Others have invoked  \ion{C}{2} $\lambda$6580 blended 
 with H$\alpha$ \citep{deng00} or only  \ion{Ne}{1}  $\lambda$6402 \citep{benetti02}. Recently, more detailed calculations reported by  \citet{ketchum08} suggest \ion{Si}{2} blended with \ion{Fe}{2} as a viable source.
In the case of SN~2007Y we lean towards an identification of H$\alpha$ possibly 
blended with \ion{Si}{2}.
The  hydrogen hypothesis is supported in part by the H$\alpha$ emission in the nebular spectrum
that, similar to early epochs, is at a higher velocity than the other material.
Additional evidence for hydrogen comes 
 from a notch in the earliest spectra (see Fig.~10) which may be due to
  H$\beta$ blue-shifted by $\sim$14,000 km s$^{-1}$.
With these pieces of evidence, and the arguments laid out by \citet{anupama05}, \citet{folatelli06} and \citet{parrent07} for  the case of SN~2005bf, we conclude 
that H$\alpha$ is likely the dominant ion responsible for the absorption at 6200 \AA.
The gross inconsistency between the blue-shift derived from
\ion{Si}{2}, compared to the other ions (see Fig.~\ref{fig:velocities}),
 allows us to reject \ion{Si}{2} as the only ion producing this feature. 
To decisively ascertain to what extent these or other ions attribute 
to the 6200 \AA\ feature remains to be determined through
sophisticated radiative transfer calculations.
  
  The identification of high-velocity calcium gas has also 
  been reported in the early phase spectra of SN~1999ex and SN~2005bf. 
  The  strong spectroscopic similarity of these events to SN~2007Y, and the fact 
  that by maximum light both SN~1999ex and SN~2007Y are representative of a normal  Type~Ib event are both suggestive that this may be a generic feature of many SNe~Ib.
    The fast evolution of the \ion{Ca}{2} triplet highlights the 
  importance to obtain pre-maximum spectra of these events in order to confirm if 
  high-velocity calcium gas is in fact a commonality among other SNe~Ib/c. 
  Note, however, there appears to be no signature of 
  this feature in the early epoch spectra of SN~2008D.

 We estimate 0.06~M$_{\sun}$ of $^{56}$Ni was synthesized 
 in the explosion.
 We also find from modeling the nebular spectra a lower limit on the ejected oxygen mass below 4500 km s$^{-1}$ of 0.20 M$_{\sun}$, and a total ejecta mass of 0.44 M$_{\sun}$. Stellar evolutionary models predict this amount of oxygen is produced 
 from He stars with a mass of $\sim$3.3~M$_{\sun}$ 
  \citep[][]{thielemann96,woosley95,shigeyama90}.  The He star most likely
  evolved from a ZAMS star of $\approx$10--13~M$_{\sun}$
  that formed in a binary system, and during its evolution, was able to shed its 
  envelope through stable case C mass transfer \citep[for an example see ][]{pods93}. 
  In this scenario, prior to explosion, the He star retains a residual hydrogen shell 
  with a mass of a few tenths of a solar mass or less. 
  Alternatively, but more unlikely given the abundance pattern, the progenitor star could have been a massive star (M$_{\rm ZAMS}$ $>$ 35 M$_{\sun}$) that underwent strong stellar winds prior to explosion \citep{woosley93}. 
A scenario of a more massive star is also disfavored given the lack of evidence for
asymmetry  in the ejecta.
  
  Now we attempt to estimate the kinetic energy of the explosion.
  As previously discussed in $\S$~\ref{lightcurves},
  the similarity between the light curve shape of 
  SNe~2006aj and 2007Y  implies M$_{\rm ej}$(07Y)$^3$ / E$_{\rm KE}$(07Y)  $\propto$ 
 M$_{\rm ej}$(06aj)$^3$ / E$_{\rm KE}$(06aj). Plugging in M$_{\rm ej}$(06aj) $=$ 1.8 
 M$_{\sun}$, E$_{\rm KE}$(06aj) $=$ 2.0$\times10^{51}$ erg \citep{maeda07a,mazzali06b}
 and  M$_{\rm ej}$(07Y) $=$ 0.50 M$_{\sun}$, and then solving for E$_{\rm KE}$(07Y),
 gives $\approx$4.3$\times$10$^{49}$ erg.
 Alternatively, simply looking at the ejecta mass and the velocity of the nebula 
 E$_{\rm KE}$(07Y) $\approx$  $(1/2)$$\cdot$M$_{\rm ej}$(07Y)$\cdot$v$_{neb}^{2}$ 
 $\approx$ 1.0$\times$10$^{50}$ erg.
 
 A more accurate approach is to combine the fact that the light curve shapes of SN~2007Y and SN~2006aj are similar to the ratio of their photospheric expansion velocity via 
 v$_{\rm ph}$$\propto$(E$_{\rm KE}$/M$_{\rm ej}$)$^{1/2}$.
 At maximum light the ratio of the photospheric velocity of SN~2007Y (v$_{\rm ph} = 9000$ km~s$^{-1}$) to SN~2006aj (v$_{\rm ph} = 15000$ km~s$^{-1}$; Pian et al. 2006)  is 0.60. This implies a E$_{\rm KE}$(07Y) $\sim$ 1.8$\times$10$^{50}$ erg. If we double 
 M$_{\rm ej}$(07Y) then   E$_{\rm KE}$(07Y)  increases by a factor of two. A
E$_{\rm KE}$(07Y) of $\approx$few$\times$10$^{50}$ erg is nearly an order of 
magnitude less 
than the $\geq$10$^{51}$ erg that is found in other SNe~Ib/c.
 
In Fig.~\ref{fig:parameters} the physical parameters of SN~2007Y are compared to those of 13  other well-studied core-collapse SNe.
Here  the estimated values of M$_{\rm ZAMS}$ are compared to (top panel) E$_{\rm KE}$
and (bottom panel) the ejected $^{56}$Ni mass. From its position in the top panel we 
find SN~2007Y to be one of the least energetic SNe~Ib/c yet studied.
Correspondingly, the $^{56}$Ni mass is comparable to the least luminous SNe~Ib/c 
while the M$_{\rm ZAMS}$ lies at the low end of the M$_{\rm ZAMS}$ distribution 
of the comparison sample, most similar to SN~1993J.

 \section{Conclusion}
 We present detailed photometric and spectroscopic observations of SN~2007Y.  
 SN~2007Y appears to be one of the least energetic normal SN~Ib yet studied, 
 with E$_{\rm KE}\sim$10$^{50}$ erg. We estimate  a total 
 ejected mass of 0.45 M$_{\sun}$, of which 
 0.06 M$_{\sun}$ is $^{56}$Ni and 0.2 M$_{\sun}$ is O.
 The oxygen abundance suggests the progenitor was most likely a 
3.3 M$_{\sun}$  He-core star that evolved from a 10--13 M$_{\sun}$ main-sequence mass 
star in a binary system. These physical parameters all point towards a progenitor 
similar to that of SN~1993J.
 
 The remarkable similarity of the  early phase spectra
of SN~2007Y compared to  SN~1999ex and SN~2005bf, suggests these
events had very similar photospheric conditions. However,
subtle difference do exist. These are most likely related to differences 
in the total ejected mass and the configuration of the ejecta. 
In the earliest phases we identify high-velocity features of H$\alpha$ and \ion{Ca}{2}.
These features maybe a commonality in many other SNe~Ib/c. 
This remains to be confirmed with early phase observations  
that are expected to be performed in future all-sky surveys. 
  

\begin{appendix}
\section{Observations and Data Reduction}
\subsection{Photometry}

The UVOT light curves were originally presented in  \citet{brown08},
and the reader is referred to that paper and references therein for details of the
data acquisition and reduction process. The light curves have superb sampling that 
cover the evolution of SN~2007Y from $-$11.6 to 21.6 days past maximum light.

 Twenty-one epochs of early-time optical ($u'g'r'i'BV$) 
 imaging were  acquired with the  LCO's Henrietta Swope 1-m telescope equipped with a 
 direct imaging CCD camera, and two epochs of spectroscopic acquisition imaging  
 with the du~Pont ($+$WFCCD:  Wide Field Re-imaging CCD Camera) 2.5-m telescope. 
 We direct the reader to \citet[][Paper I]{hamuy06} and \citet{contreras09} for a detailed technical description of the telescopes, instruments, and filter transmission functions used in the low-z portion of  the CSP. 
 Paper~I also contains a step by step account of the data reduction process for each instrument used by the  CSP. 
 
The brightness of SN~2007Y was determined differentially with respect to a 
sequence of local field stars. 
Absolute photometry of the local sequence
in the optical (Table~1) was obtained using 
standard fields \citep{landolt92,smith02} observed on 4 photometric nights. 

Instrumental magnitudes of the standard fields were computed using the 
IRAF\footnote[25]{The image Reduction and Analysis Facility (IRAF) is 
maintained and distributed by the Association of Universities for Research
in Astronomy, under the cooperative agreement with the National Science 
Foundation.}  
 {\tt DAOPHOT} package {\tt PHOT} with an aperture radius of 7$\arcsec$. 
Adopting extinction coefficients and color terms derived through the course 
 of the CSP (see Fig.~3  and Fig.~4 in Paper I)
the instrumental magnitudes of the standard stars were used to derive
zero-points to calibrate the local sequence.


The magnitude of SN~2007Y and the local sequence were measured by a point-spread function (PSF) technique.
A PSF within a radius of 3$\arcsec$ was fitted to each sequence star and the supernova.
A night-to-night zero-point was computed from the local sequence and then
used to obtain the apparent magnitude of SN~2007Y.

Although the supernova was located well outside the spiral arms, close inspection of 
optical VLT images obtained under excellent seeing conditions 
revealed the presence of an   
 \ion{H}{2} region adjacent to the location of SN~2007Y. To assess if  our photometry was contaminated by either this \ion{H}{2} region
or any spatial/time varying background, optical galaxy templates of  NGC~1187 were 
obtained on March 5.2, 
2008 with a direct imaging camera (Tek~5) attached to the du~Pont. 
These templates were used to subtract any background on all early-time CSP 
$u'g'r'i'BV$ images. We found mean magnitude 
differences for each light curve 
(in the sense of photometry computed with versus without template subtraction) of  
$\Delta$($u',g',r',i,B,V) =  (0.009, 0.003, 0.004, 0.001, 0.003, 0.003)$.
These small differences indicate that no systematic offset was introduced to the images
from template subtraction and that any background in the optical is negligible.


Two epochs of late-time $BVRI$ imaging at 270 and 343 days past L$_{\rm max}$
were obtained with the VLT (+FORS1) in service mode.
Science product images reduced by the ESO pipeline 
were used for the aperture photometry. As all of the local sequence 
stars used in the early phase imaging were saturated in the VLT images it proved 
necessary to define another local sequence. Absolute  photometry of this 
local sequence was obtained using the standard star field PG2331+055 
\citep{landolt92} observed during the first photometric epoch of VLT imaging.
Instrumental magnitudes of the 3 stars in PG2331+055 were computed using
{\tt PHOT} with an aperture radius of 7$\arcsec$.
These instrument magnitudes were then corrected for atmospheric extinction
and color terms using standard values derived for  Chip 1 of FORS1.\footnote[26]{http://www.eso.org/observing/dfo/quality/FORS1/qc/photcoeff/photcoeffs\_fors1.html} 
 Zero-points were then computed by comparing the instrumental magnitudes to the Landolt standard magnitudes.
Instrumental magnitudes of 7 local sequence stars (and SN~2007Y) were then 
computed with a 0$\farcs$5 aperture radii plus an aperture correction.
After color and 
atmospheric extinction corrections the zero-points derived from PG2331+055 were 
applied to these magnitudes to calibrate the VLT local sequence.  This local sequence was 
then used to determine nightly zero-points in order to obtain apparent magnitudes
of SN~2007Y. 
 
Fifteen epochs of early phase near-infrared photometry 
were acquired at LCO with the Swope ($+$RetroCam) and 
du~Pont ($+$WIRC:  Wide field InfraRed Camera) telescopes. 
The majority of $YJH$-band imaging was taken with RetroCam  while
several epochs including $K_{s}$ band were obtained with WIRC.

Imaging was generally performed with a jitter technique that consists of a two loops of 
9 exposures.
Each of the individual images were dark-subtracted, flat-fielded  
and then stacked to produce a master image used to calculate photometry.

Instrumental magnitudes of standard star fields \citep{persson98} observed 
over 5 photometric nights were measured with an aperture radius
of 5$\arcsec$. Zero-points were then derived from these observations in order to 
calibrate a local sequence whose instrumental magnitudes were computed via
a small aperture and aperture correction. Standard extinction coefficients  \citep{persson98}
were used and no color term corrections were applied. In the K$_{s}$-band photometry 
we were unable to calibrate a local sequence due to a lack of standard field observations. 
Fortunately  4 of the stars that make up our local sequence were also observed by the 2MASS survey. 
We therefore adopted 2MASS magnitudes for these stars and the corresponding values are 
reported in Table~1. 
Finally, differential photometry
of SN~2007Y  (Table 3) was obtained relative to the local sequence. 

Late epoch imaging was performed with the Infrared Spectrometer And Array Camera (ISAAC) attached to the VLT. Imaging was performed in the short wavelength mode with
multiple loops of the jitter--offset mode.  Reduction of this data  was 
performed with the {\tt Eclipse} software package \citep{devillard97}. 
The  {\tt jitter} program was used to estimate and remove the sky background from each
on source image. These were then combined to produce final stacked images that were used to derive instrumental magnitudes. Absolute photometry was determined using 
several Persson standard star fields observed over the course of the night the supernova was monitored. 
We were only able to measure the brightness of SN~2007Y  
in the $J_{\rm S}$- and $H$-band in the first epoch of near-infrared 
imaging i.e. day $+$268. 
In the cases when the supernova was not detected a 3$\sigma$ upper limit of
its brightness was derived.

\subsection{Spectroscopy}

Twelve epochs of long slit spectroscopy was obtained from 
$-$14 to $+$270 days past L$_{\rm max}$. 
All spectra were reduced in the same manner using standard techniques as described
in Paper~I. In summary, each spectrum was overscan subtracted, bias corrected, 
trimmed, flat-fielded and then extracted. An accurate wavelength calibration solution was derived from arc lamp spectra and applied to the extracted one-dimensional spectrum. 
Finally, a nightly response function derived from standard star observations 
was used to flux-calibrate each spectrum.  When multiple spectra were obtained 
for a given epoch they were combined to produce a master spectrum.

\end{appendix}

\acknowledgments 
This material is based upon work supported by the National Science
Foundation (NSF) under grant AST--0306969. 
The Dark Cosmology Centre is funded by the Danish NSF.
M.S. acknowledges support from the MPA's visitor programme.
J.S. is a Royal Swedish Academy of Sciences Research Fellow supported by a
grant from the Knut and Alice Wallenberg Foundation.
MH ackowledges support from Iniciativa Cientifica Milenio
through grant P06-045-F and CONICYT through Centro de Astrofisica
FONDAP (grant 15010003), Programa Financiamiento Basal (grant PFB 06),
and Fondecyt (grant 1060808).
We have made use of the NASA/IPAC Extragalactic
Database (NED) which is operated by the Jet Propulsion Laboratory,
California Institute of Technology, under contract with the National
Aeronautics and Space Administration. This publication makes use of data products from the Two Micron All Sky Survey, which is a joint project of the University of Massachusetts and the Infrared Processing and Analysis Center/California Institute of Technology, funded by the National Aeronautics and Space Administration and the National Science Foundation. 

\clearpage
\begin{landscape}
\begin{deluxetable} {lcccccccccccc}
\tabletypesize{\scriptsize}
\tablenum{1}
\tablewidth{0pt}
\tablecaption{Photometry of the local sequence stars in the field of NGC~1187\label{tab:optirstds}}
\tablehead{
\colhead{Star} & 
\colhead{} & 
\colhead{}   &
\colhead{}   &
\colhead{}   & 
\colhead{}   & 
\colhead{}   & 
\colhead{}   &
\colhead{}   &
\colhead{}   &
\colhead{}   &
\colhead{}   &
\colhead{}   \\
\colhead{ID} &
\colhead{RA}   &
\colhead{DEC}   &
\colhead{$u'$} &
\colhead{$g'$} & 
\colhead{$r'$} & 
\colhead{$i'$} & 
\colhead{$B$} & 
\colhead{$V$} &
\colhead{$Y$} &
\colhead{$J$} &
\colhead{$H$} &
\colhead{$K_s$} }
\startdata
C1  & 45.72981 & -22.85382 & 19.068(054) &  16.163(015) &   $\cdots$   &  $\cdots$    &  16.881(013) &  $\cdots$    &   $\cdots$   &  $\cdots$    &  $\cdots$ &  $\cdots$ \\
C2  & 45.59832 & -22.83040 & 17.267(011) &  15.961(007) &  15.449(009) &  15.244(009) &  16.362(009) &  15.668(008) &   $\cdots$   &  $\cdots$    &  $\cdots$ &  $\cdots$ \\
C3  & 45.69047 & -22.82723 & 17.329(012) &  16.114(007) &  15.683(008) &  15.520(008) &  16.489(011) &  15.861(007) &   $\cdots$   &  $\cdots$    &  $\cdots$ &  $\cdots$ \\
C4  & 45.65395 & -22.83600 & 19.246(077) &  16.630(013) &  15.415(008) &  14.873(007) &  17.292(015) &  15.996(011) &   $\cdots$   &  $\cdots$    &  $\cdots$ &  $\cdots$ \\
C5  & 45.60573 & -22.93198 & 17.840(014) &  16.669(007) &  16.254(009) &  16.093(009) &  17.023(016) &  16.422(007) &   $\cdots$   &  $\cdots$    &  $\cdots$ &  $\cdots$ \\
C6  & 45.67210 & -22.84998 & 19.925(092) &  17.192(007) &  15.967(010) &  15.440(007) &  17.878(014) &  16.547(012) &   $\cdots$   &  $\cdots$    &  $\cdots$ &  $\cdots$ \\
C7  & 45.61845 & -22.88566 & 18.495(024) &  16.989(012) &  16.450(011) &  16.256(007) &  17.419(019) &  16.683(012) &   $\cdots$   &  $\cdots$    &  $\cdots$ &  $\cdots$ \\
C8  & 45.59823 & -22.86743 & 18.666(033) &  17.391(007) &  16.907(007) &  16.725(009) &  17.783(017) &  17.099(010) &   $\cdots$   &  $\cdots$    &  $\cdots$ &  $\cdots$ \\
C9  & 45.58725 & -22.93928 & 19.183(091) &  17.521(011) &  16.859(033) &  16.571(011) &  18.025(031) &  17.131(025) &   $\cdots$   &  $\cdots$    &  $\cdots$ &  $\cdots$ \\
C10 & 45.64463 & -22.84426 & 19.173(047) &  17.583(008) &  16.966(009) &  16.720(012) &  18.053(019) &  17.233(018) &   $\cdots$   &  $\cdots$    &  $\cdots$ &  $\cdots$ \\
C11 & 45.61671 & -22.85780 & 19.763(087) &  17.742(015) &  16.944(010) &  16.643(007) &  18.251(014) &  17.307(019) &   $\cdots$   &  $\cdots$    &  $\cdots$ &  $\cdots$ \\
C12 & 45.62911 & -22.84027 & 18.581(028) &  17.687(008) &  17.385(008) &  17.283(011) &  17.975(016) &  17.514(009) &   $\cdots$   &  $\cdots$    &  $\cdots$ &  $\cdots$ \\
C13 & 45.67855 & -22.85963 & 19.859(086) &  17.862(009) &  17.145(009) &  16.857(010) &  18.353(018) &  17.483(008) &   $\cdots$   &  $\cdots$    &  $\cdots$ &  $\cdots$ \\
C14 & 45.62110 & -22.94564 & 20.726(185) &  18.239(014) &  16.897(011) &  16.162(008) &  19.012(026) &  17.530(010) & 15.018(027) &   14.612(017) &   13.996(034) &  $\cdots$\\
C15 & 45.63166 & -22.92610 & 20.736(193) &  18.483(029) &  17.140(013) &  16.445(007) &  19.128(058) &  17.757(011)& 15.334(017) &   14.939(013) &   14.328(050) &   14.157(059) \\
C16 & 45.59841 & -22.85483 & 20.471(150) &  18.473(017) &  17.810(008) &  17.563(009) &  18.983(025) &  18.125(011) &   $\cdots$   &  $\cdots$    &  $\cdots$ &  $\cdots$ \\
C17 & 45.59876 & -22.91843 & 21.148(362) &  18.887(042) &  17.529(009) &  16.738(014) &  19.622(044) &  18.141(016) &   $\cdots$   &  $\cdots$    &  $\cdots$ &  $\cdots$ \\
C18 & 45.58554 & -22.85900 & 21.906(830) &  18.935(034) &  17.621(018) &  16.868(011) &  19.543(115) &  18.235(046) &   $\cdots$   &  $\cdots$    &  $\cdots$ &  $\cdots$ \\
C19 & 45.67666 & -22.91654 & 21.821(535) &  18.908(025) &  17.592(007) &  16.481(007) &  19.708(048) &  18.194(015) & 15.099(016) &   14.611(022) &   14.087(040) &   13.867(063)\\
C20 & 45.62264 & -22.92669 & 22.066(983) &  19.108(030) &  17.774(009) &  16.481(007) &  20.005(068) &  18.353(013)& 14.993(050) &   14.527(013) &   14.001(052) &   13.685(048) \\
C21 & 45.61848 & -22.88557 & $\cdots$   &  $\cdots$    &  $\cdots$ &  $\cdots$ &  $\cdots$    &   $\cdots$ & 15.532(063) &   15.281(012) &   14.954(048) &   14.811(122) \\
C22 & 45.73816 & -22.90095 & $\cdots$   &  $\cdots$    &  $\cdots$ &  $\cdots$ &  $\cdots$    &   $\cdots$ & 16.023(026) &   15.605(015) &   15.038(067) &  $\cdots$ \\
C23 & 45.70857 & -22.96788 & $\cdots$   &  $\cdots$    &  $\cdots$ &  $\cdots$ &  $\cdots$    &   $\cdots$ & 16.434(049) &   16.002(023) &   15.547(033) &  $\cdots$ \\
C24 & 45.71329 & -22.93300 & $\cdots$   &  $\cdots$    &  $\cdots$ &  $\cdots$ &  $\cdots$    &   $\cdots$ & 16.930(041) &   16.321(035) &   15.848(062) &  $\cdots$ \\
C25 & 45.73165 & -22.92826 & $\cdots$   &  $\cdots$    &  $\cdots$ &  $\cdots$ &  $\cdots$    &   $\cdots$ & 16.641(033) &   16.146(030) &   15.629(054) &  $\cdots$ \\
C26 & 45.71723 & -22.92116 & $\cdots$   &  $\cdots$    &  $\cdots$ &  $\cdots$ &  $\cdots$    &   $\cdots$ & 16.803(076) &   16.322(038) &   15.844(064) &  $\cdots$ \\
C27 & 45.70581 & -22.95958 & $\cdots$   &  $\cdots$    &  $\cdots$ &  $\cdots$ &  $\cdots$    &   $\cdots$ & 16.961(051) &   16.522(036) &   16.065(062) &  $\cdots$ \\
C28 & 45.63538 & -22.93916 & $\cdots$   &  $\cdots$    &  $\cdots$ &  $\cdots$ &  $\cdots$    &   $\cdots$ & 17.184(065) &   16.611(079) &   16.317(094) &  $\cdots$ \\
C29 & 45.67655 & -22.86069 & $\cdots$   &  $\cdots$    &  $\cdots$ &  $\cdots$ &  $\cdots$    &   $\cdots$ & 15.303(023) &   14.915(022) &   14.278(027) &  $\cdots$ \\

\enddata
\tablecomments{Equinox of the coordinates is 2000.0. Uncertainties given in parentheses in millimag
  correspond to the rms of the magnitudes obtained on four and five
  photometric nights in the optical and near-infrared, respectively.
  The magnitudes of the four $K_{s}$-band stars
  are the standard values from the 2mass catalog.}
\end{deluxetable}

\clearpage
\end{landscape}

\begin{deluxetable} {lccccccc}
\tabletypesize{\scriptsize}
\tablecolumns{8}
\tablenum{2}
\tablecaption{CSP optical photometry of SN~2007Y\label{tab:optphot}}
\tablehead{
\colhead{JD$-2,400,000$} &
\colhead{$u'$} & 
\colhead{$g'$} & 
\colhead{$r'$} & 
\colhead{$i'$} & 
\colhead{$B$} & 
\colhead{$V$} &
\colhead{Instrument}} 
\startdata      
54150.53& $ \cdots $ & $ \cdots $ & $ \cdots $ & $ \cdots $ & 17.658(024)& 16.994(024)& Tek  5\\
54153.51& $ \cdots $ & $ \cdots $ & $ \cdots $ & $ \cdots $ & 16.724(015)& 16.386(015)& Site 3\\
54154.51& 16.990(015)& 16.275(015)& 16.222(015)& 16.336(015)& 16.481(015)& 16.205(015)& Site 3\\
54155.52& $ \cdots $ & $ \cdots $ & 16.075(015)& 16.184(015)& 16.282(015)& 16.054(015)& Site 3\\
54156.51     & $\cdots$    &   $\cdots$    & $\cdots$    & $\cdots$     & 16.210(064) &  $\cdots$    & Tek  5\\
53158.52& 16.064(015)& 15.700(015)& 15.722(015)& 15.804(015)& 15.864(015)& 15.687(015)& Site 3\\
53159.52& 15.960(015)& 15.608(015)& 15.634(015)& 15.717(015)& 15.775(015)& 15.596(015)& Site 3\\
53160.51& 15.877(016)& 15.552(015)& 15.553(015)& 15.620(015)& 15.717(015)& 15.526(015)& Site 3\\
53161.50& 15.845(021)& 15.506(015)& 15.487(015)& 15.552(015)& 15.692(015)& 15.469(015)& Site 3\\
53163.51& 15.921(015)& 15.432(015)& 15.364(015)& 15.423(015)& 15.644(015)& 15.372(015)& Site 3\\
53165.50& 16.183(016)& 15.461(015)& 15.297(015)& 15.336(015)& 15.707(015)& 15.332(015)& Site 3\\
53166.52& 16.452(015)& 15.518(015)& 15.286(015)& 15.325(015)& 15.789(015)& 15.349(015)& Site 3\\
53169.50& 17.398(015)& 15.800(015)& 15.370(015)& 15.382(015)& 16.188(015)& 15.519(015)& Site 3\\
53170.50& 17.716(016)& 15.923(015)& 15.442(015)& 15.408(015)& 16.349(015)& 15.620(015)& Site 3\\
53171.51& $ \cdots $ & $ \cdots $ & 15.517(015)& 15.444(015)& 16.520(015)& 15.719(015)& Site 3\\
53174.50& 18.887(045)& 16.431(015)& 15.727(015)& 15.565(015)& 16.979(015)& 16.014(015)& Site 3\\
53179.49& 19.420(081)& 16.858(015)& 16.005(015)& 15.741(015)& 17.492(015)& 16.354(015)& Site 3\\
53185.50& 19.842(157)& 17.147(015)& 16.246(015)& 15.981(015)& 17.816(015)& 16.623(015)& Site 3\\
53190.49& 19.740(213)& 17.278(015)& 16.404(015)& 16.127(015)& 17.915(021)& 16.726(015)& Site 3\\
53194.48& 19.418(146)& 17.384(015)& 16.514(015)& 16.231(015)& 17.956(026)& 16.845(015)& Site 3\\
53197.48& 19.478(103)& 17.429(015)& $ \cdots $ & $ \cdots $ & 18.063(015)& $ \cdots $ & Site 3\\
54204.47& 19.399(152)& $ \cdots $ & $ \cdots $ & $ \cdots $ & $ \cdots $ & 17.034(015)& Site 3\\
54204.48& 19.408(104)& $ \cdots $ & $ \cdots $ & $ \cdots $ & $ \cdots $ & $ \cdots $ & Site 3\\
54434.11     & $\cdots$    &   $\cdots$    & 20.951(077) & 20.658(108)  & 22.517(127) &  21.900(081) & FORS1 \\
54507.03& $\cdots$    &   $\cdots$    & 22.151(162) & 21.983(150)  & 23.521(185) &  23.159(125) & FORS1 \\
\enddata
\tablecomments{Bolometric maximum occured on JD-2454163.12. 
Uncertainties given in parentheses are in millimag. 
FORS1 imaging was conducted with a set of standard Johnson Kron-Cousins $BVRI$ filters.}
\end{deluxetable}

\begin{deluxetable}{lccccc}
\tabletypesize{\normalsize}
\tablecolumns{6}
\tablenum{3}
\tablecaption{CSP near-infrared photometry of SN~2007Y\label{tab:irphot}}
\tablehead{
\colhead{JD$-2,400,000$} &
\colhead{$Y$} & 
\colhead{$J$} & 
\colhead{$H$} & 
\colhead{$K_{s}$} & 
\colhead{Instrument}} 
\startdata      
54150.51      & 16.679(036) &   16.628(044) &  16.618(063) &   $\cdots$    &    RetroCam  \\
54156.53     & 15.631(015) &   15.628(014) &  $\cdots$    &   $\cdots$    &    RetroCam  \\
54158.54      & 15.416(012) &   15.397(012) &  15.341(031) &   15.187(044) &    WIRC    \\
54159.52      & 15.337(012) &   15.319(013) &  15.247(031) &   15.079(044) &    WIRC    \\
54162.50      & 15.109(015) &   15.063(013) &  14.975(028) &   $\cdots$    &    RetroCam \\ 
54164.50      & 14.995(013) &   14.946(015) &  14.864(025) &   $\cdots$    &    RetroCam \\
54167.51      & 14.935(012) &   14.830(011) &  14.748(022) &   $\cdots$    &    RetroCam  \\
54168.52      & 14.962(013) &   14.845(014) &  14.744(023) &   $\cdots$    &    RetroCam \\
54169.52      & 14.969(016) &   14.835(009) &  14.751(029) &   $\cdots$    &    WIRC     \\
54170.51      & 15.009(016) &   14.822(010) &  14.698(030) &   14.505(037) &    WIRC   \\
54171.50      & 15.031(016) &   14.888(008) &  14.743(029) &   14.481(036) &    WIRC   \\
54172.50      & 15.070(016) &   14.899(009) &  14.770(029) &   14.504(036) &    WIRC   \\
54175.51     & 15.125(014) &   15.030(019) &  14.745(031) &   $\cdots$    &    RetroCam \\
54189.50     & 15.276(021) &   15.521(027) &  15.124(049) &   $\cdots$    &    RetroCam \\
54193.49      & 15.328(016) &   15.663(025) &  15.313(040) &   $\cdots$    &    RetroCam \\
54432.26     &  $\cdots$   &   20.906(121) &  20.393(126) &   $>$21.241   &    ISAAC \\
54507.06      &  $\cdots$   &   $>$21.502   &  $>$21.380   &   $>$21.219   &    ISAAC \\
\enddata
\tablecomments{Bolometric maximum occured on JD-2454163.12.
Uncertainties given in parentheses are in millimag.}

\end{deluxetable}

\clearpage
\begin{deluxetable} {clcclccc}
\tabletypesize{\small}
\tablecolumns{5}
\tablewidth{0pt}
\tablenum{4}
\tablecaption{Spectroscopic observations of SN~2007Y\label{tab:spec}}
\tablehead{
\colhead{Phase} &
\colhead{Epoch\tablenotemark{a}} &
\colhead{Telescope} &
\colhead{Instrument} &
\colhead{Range} &
\colhead{Resolution} &
\colhead{Exposure} \\ 
\colhead{JD$-2,400,000$} &
\colhead{(days)} &
\colhead{} &
\colhead{} &
\colhead{(\AA)} &
\colhead{(\AA\ per pixel)} &
\colhead{(sec)}}

\startdata
54150.54  &  $-$14   & Du~PONT & WFCCD  & 3800 -- 9235   & 3.0  &  600   \\
54156.52  &  $-$8    & Du~PONT & WFCCD  & 3800 -- 9235   & 3.0  &  600   \\
54163.57 &  $-$1    & NTT     & EMMI   & 3200 -- 10200  & 3.3  &  200     \\
54170.50 &  $+$6    & BAADE   & IMACS  & 3790 -- 10300  & 2.0  &  600     \\
54173.51 &  $+$9    & Du~PONT & B\&C   & 3430 -- 9650   & 3.0  &  600     \\
54178.50 &  $+$14   & Du~PONT & B\&C   & 3420 -- 9630   & 3.0  &  600     \\ 
54185.49 &  $+$21   & Du~PONT & B\&C   & 3400 -- 9600   & 3.0  &  600      \\
54203.48 &  $+$39   & Du~PONT & WFCCD  & 3800 -- 9235   & 3.0  &  800    \\
54363.82 &  $+$201  & NTT     & EMMI   & 3750 -- 8160   & 3.0 &  1800    \\
54394.86 &  $+$230  & CLAY    & LDSS3  & 4100 -- 9980   & 1.2   &  900\tablenotemark{b}     \\
54421.78 &  $+$257  & BAADE   & IMACS  & 3800 -- 10675  & 2.0  &  1800  \\
54434.57 &  $+$270  & VLT     & FORS1  & 3640 -- 8880   & 3.3  &  1280     \\
\enddata
  \tablecomments{Some spectra are the combination of multiple observations. }
\tablenotetext{a}{Days since L$_{\rm max}$.} 
\tablenotetext{b}{Different exposure times in blue channel (900 sec) and
  red channel (600 sec).}

\end{deluxetable}

\begin{deluxetable} {lccc}
\tablecolumns{4}
\tablenum{5}
\tablewidth{0pc}
\tablecaption{Lightcurve Parameters  \label{lcpar}}
\tablehead{
\colhead{Filter} &
\colhead{Peak Time} &
\colhead{Peak Obs.} &
\colhead{Peak Abs.} \\
\colhead{} &
\colhead{(JD$-2,400,000)$} &
\colhead{(Mag.)} &
\colhead{(Mag.)} }
\startdata
$w2$     & 54161.3      & 18.348(050)  & -15.236(578)\\ 
$m2$     & 54162.5      & 19.394(200)  & -15.351(616)\\
$w1$     & 54161.4      & 16.903(050)  & -17.303(578)\\
$U$      & 54161.4      & 15.057(010)  & -16.951(578)\\
$u'$     & 54161.6      & 15.849(013)  & -16.129(576)\\  
$B$      & 54163.3      & 15.647(012)  & -16.242(576)\\
$g'$     & 54163.8      & 15.437(006)  & -16.396(576)\\
$V$      & 54165.6      & 15.336(011)  & -16.446(576)\\
$r'$     & 54166.5      & 15.292(010)  & -16.430(576)\\
$i'$     & 54166.9      & 15.326(010)  & -16.325(576)\\
$Y$      & 54167.2      & 14.948(031)  & -16.597(579)\\                        
$J$      & 54169.2      & 14.839(028)  & -16.686(577)\\
$H$      & 54171.4      & 14.733(069)  & -16.757(581)\\
$K_{s}$  & 54171.4      & 14.494(640)  & -16.975(985)\\

\enddata
\tablecomments{Bolometric maximum occured on JD-2454163.12. Uncertainity in time of peak magnitude is 0.5 days for each filter
except the $m2$- and $K_{s}$-bands. Their estimated time of maximum has 
a 2 day uncertainity.
}

\end{deluxetable}

\begin{deluxetable}{lccc}
\tablecolumns{4}
\tablenum{6}
\tablewidth{0pt} 
\tablecaption{Radio Observations of SN~2007Y\label{tab:radio}}
\tablehead{
\colhead{Date Obs} & 
\colhead{$\Delta t$\tablenotemark{a}} &
\colhead{$F_{\nu,8.46~\rm GHz}$\tablenotemark{b}} &
\colhead{Array} \\
\colhead{(UT)} & 
\colhead{(days)} & 
\colhead{($\mu$Jy)} & 
\colhead{Config.} }
\startdata
2007 Feb 24.01 & 10 & $\pm 34$ & D \\
2007 Mar 10.89 & 25 & $\pm 18$ & D \\
2007 May 26.79 & 102 & $\pm 37$ & A \\
2007 Dec 21.21 & 310 & $\pm 33$ & B \\
2008 Jun 13.66 & 486 & $\pm 38$ & D \\  
2008 Dec 16.22 & 671 & $\pm 27$ & A \\
\enddata
\tablenotetext{a}{Assuming an explosion date of 2007 February 14 UT or JD-2454145.5.}  
\tablenotetext{b}{All flux densities are given as $1\sigma$ (rms).}  
\label{tab:vla}
\end{deluxetable}

\clearpage

\clearpage
\begin{figure}
\figurenum{1}
\epsscale{1.0}
\plotone{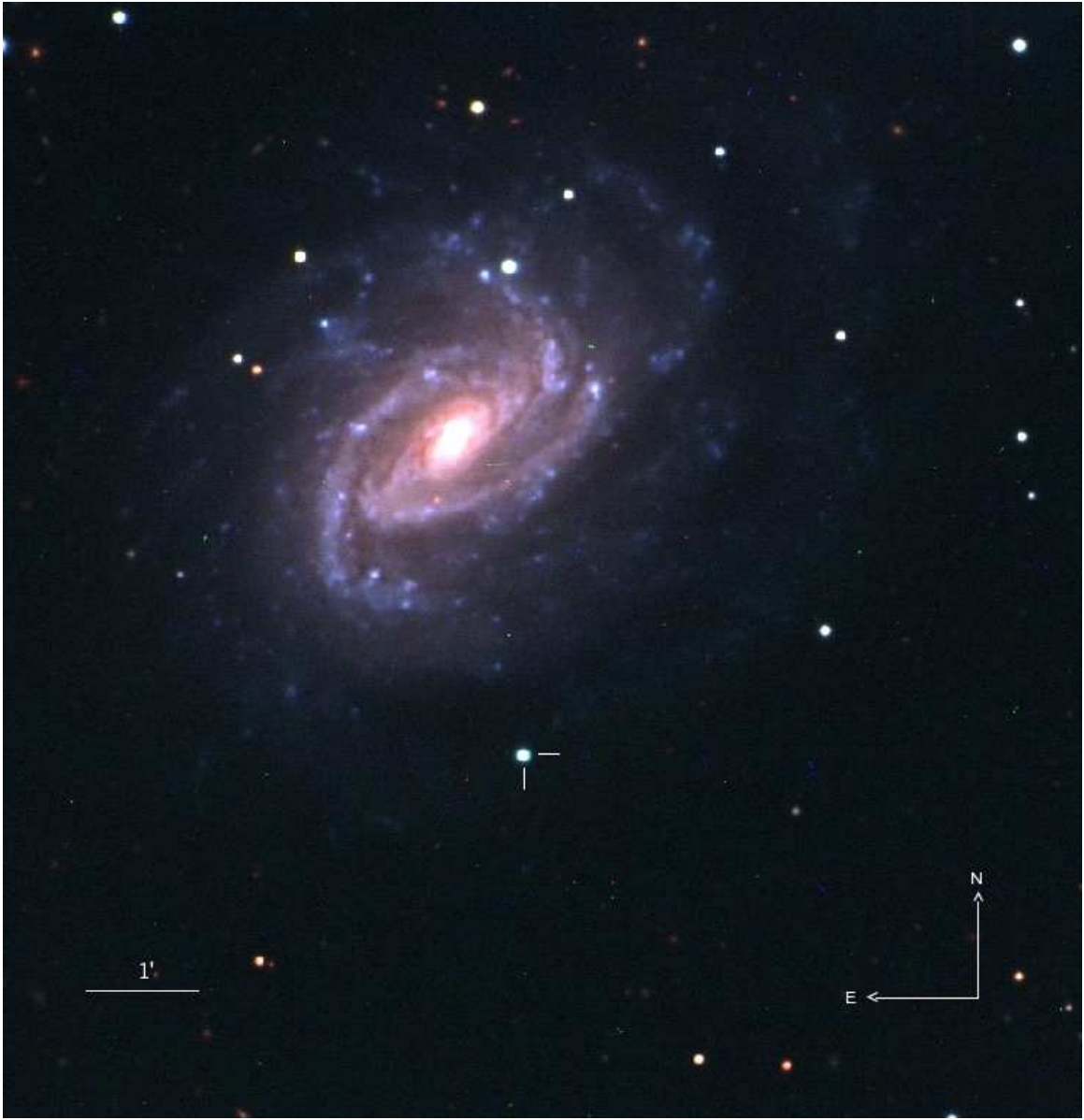}
\caption[fig1.eps]
{Color image of NGC~1187 with SN~2007Y indicated.\label{fig:FC}}
\end{figure}

\clearpage
\begin{figure}
\figurenum{2}
\epsscale{1.0}
\plotone{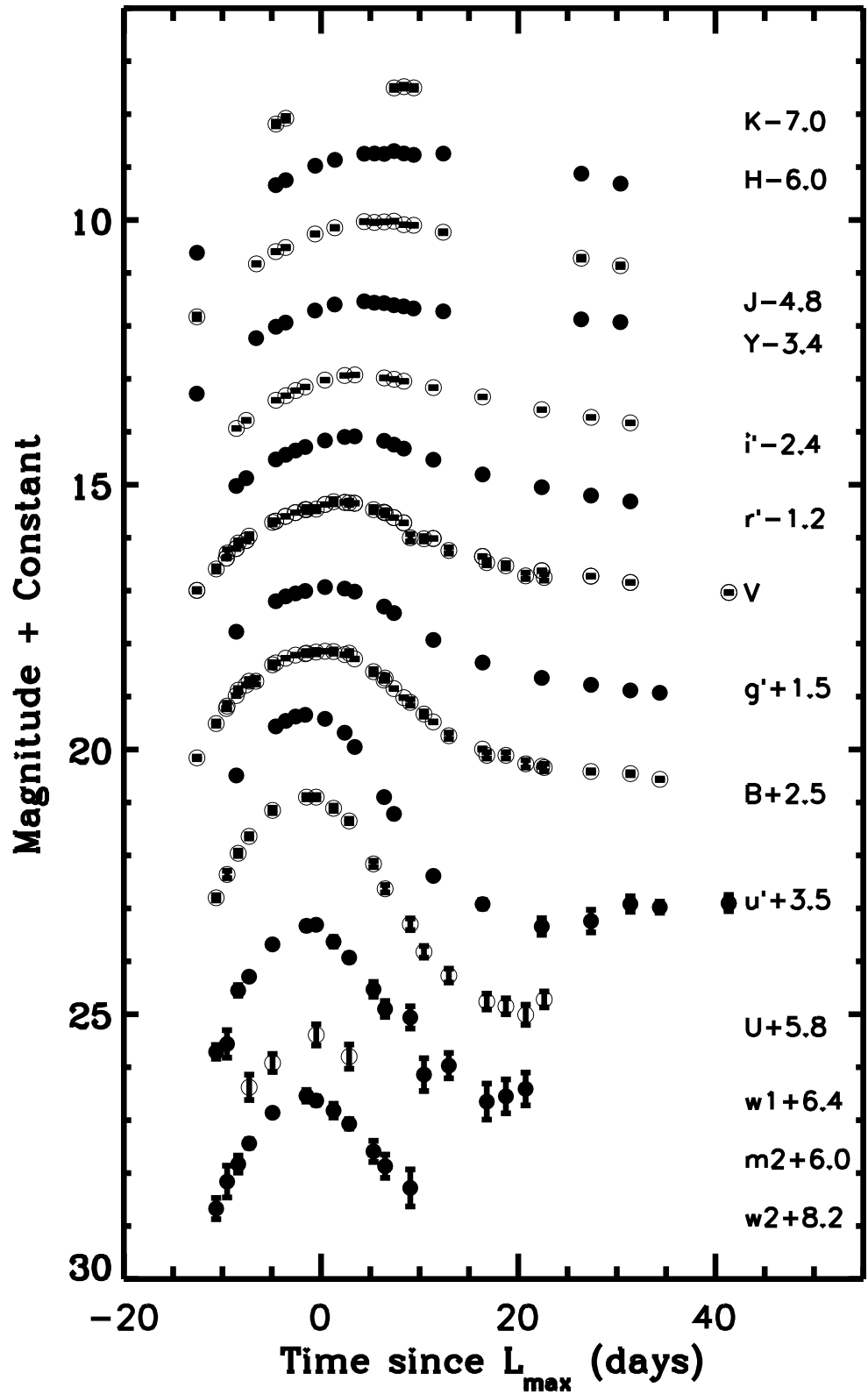}
\caption[fig2.eps]
{Observed ultraviolet, optical, and near-infrared early-time light curves of SN~2007Y plotted
as a function of time since bolometric maximum.
The light curves have been shifted in the y-direction for clarity.\label{fig:lightcurves}}
\end{figure}

\clearpage
\begin{figure}
\figurenum{3}
\epsscale{1.0}
\plotone{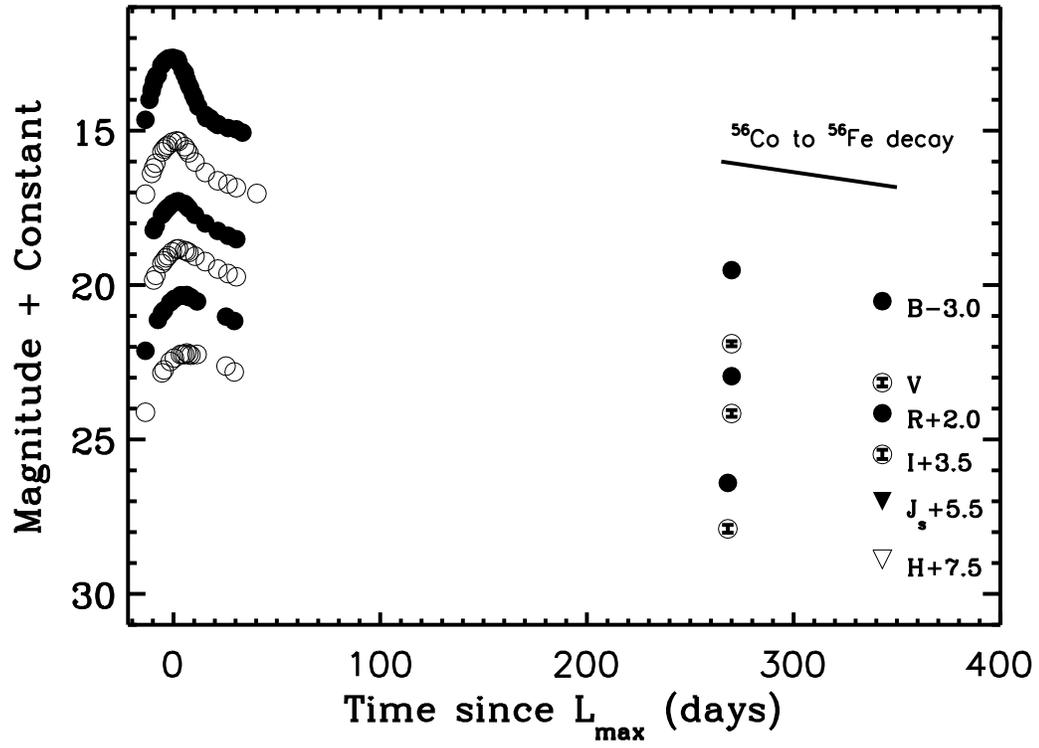}
\caption[fig3.eps]
{Late phase VLT $BVRIJ_sH$ photometry plotted with early-time
$BVr'i'JH$ light curves vs. time  since L$_{\rm max}$.
The light curves have been shifted in the y-direction for clarity as indicated.
Up side down triangles are 3$\sigma$ upper limits.\label{fig:sn07Y_LTphot}}
\end{figure}

\clearpage
\begin{figure}
\figurenum{4}
\epsscale{1.0}
\plotone{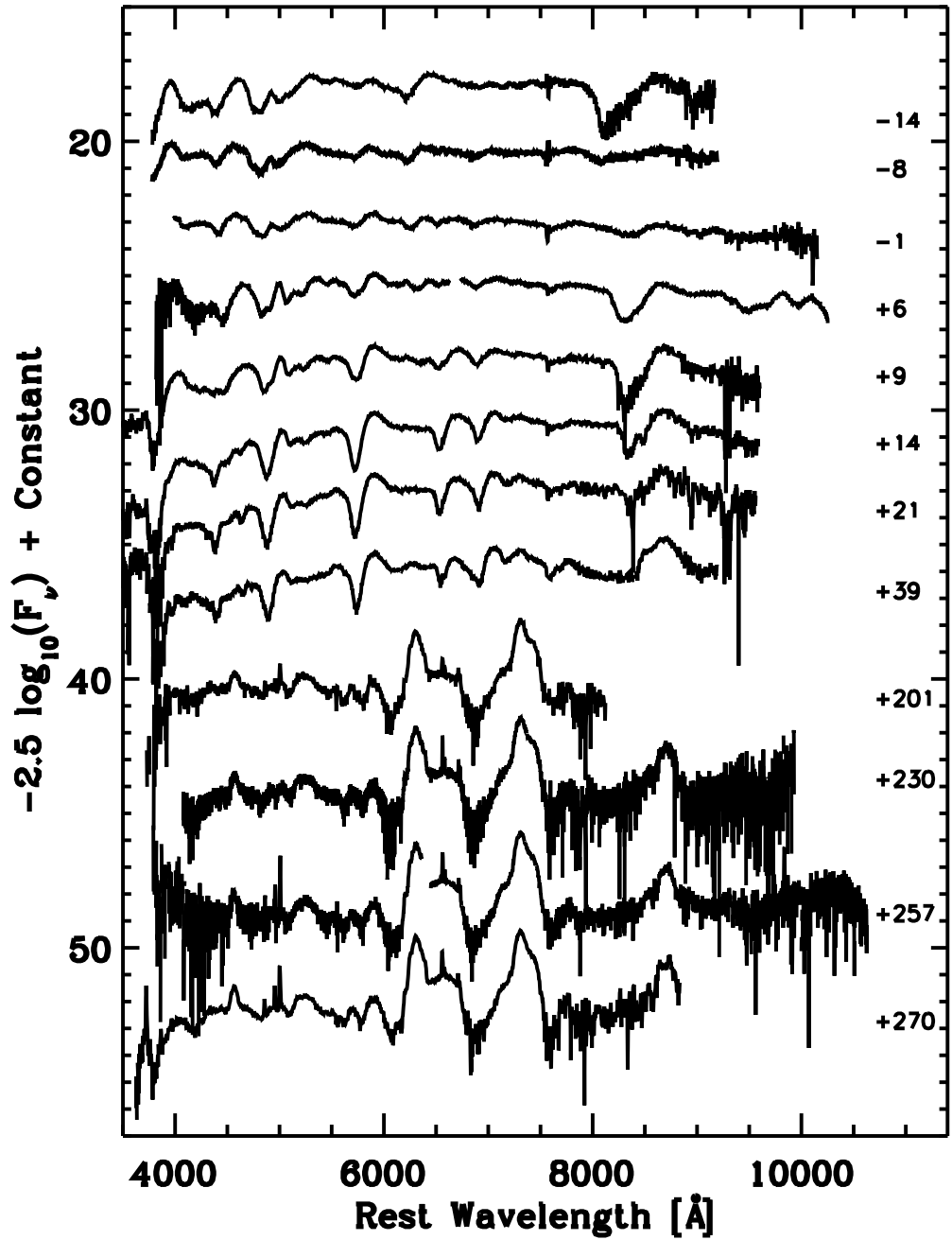}
\caption[fig4.eps]
{Spectroscopic sequence of SN~2007Y ranging from $-$14 to
$+$270 days past L$_{\rm max}$. Each spectrum is plotted
in a logarithmic (f$_{\nu}$) scale. The spectra have been shifted
below the day $-$14 spectrum by an arbitrary amount for presentation.
The labels on the right indicate the phase with respect to L$_{\rm max}$.
Each spectrum has been corrected to the rest frame of SN~2007Y adopting the
redshift z=0.00463. \label{fig:spectra}}
\end{figure}

\clearpage
\begin{figure}
\figurenum{5}
\epsscale{1.0}
\plotone{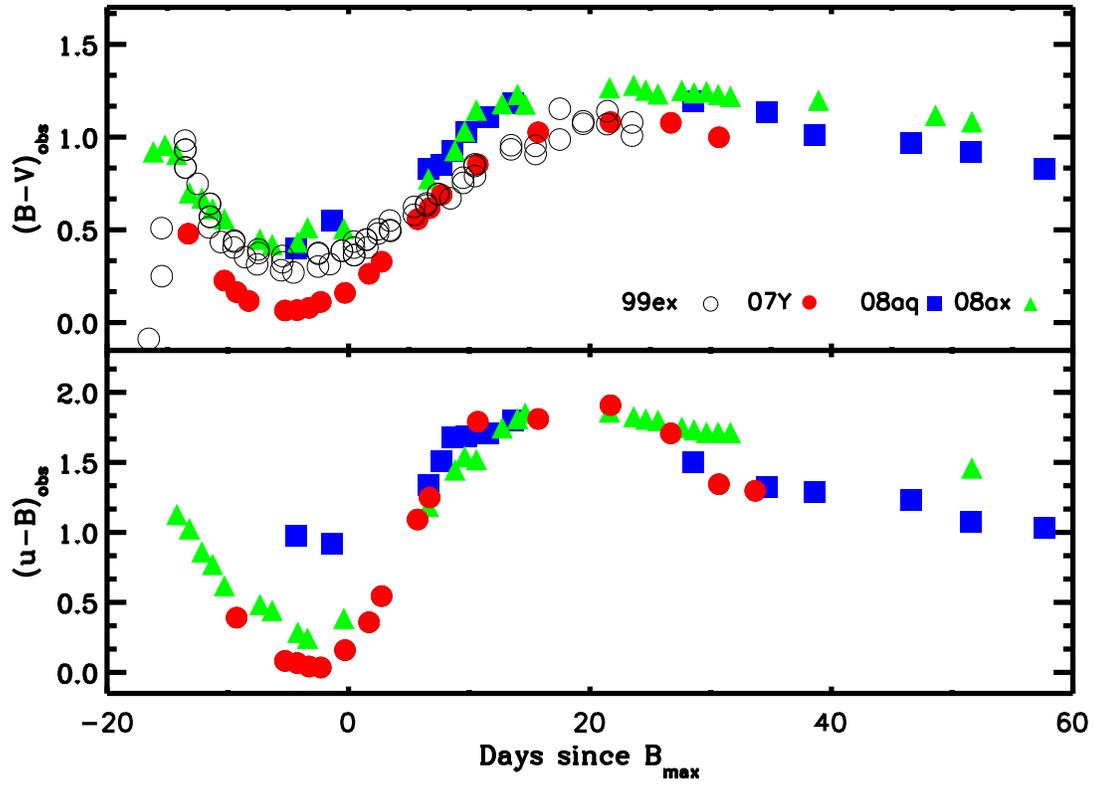}
\caption[fig5.eps]
{($u'-B$)$_{\rm obs}$ and ($B-V$)$_{\rm obs}$ color curves of SN~2007Y 
(red filled circles), SN~2008ax (green triangles), SN~2008aq (blue squares),
and the ($B-V$)$_{\rm obs}$ color curve of SN~1999ex (black circles) .\label{fig:color}}
\end{figure}

\clearpage
\begin{figure}
\figurenum{6}
\epsscale{1.0}
\plotone{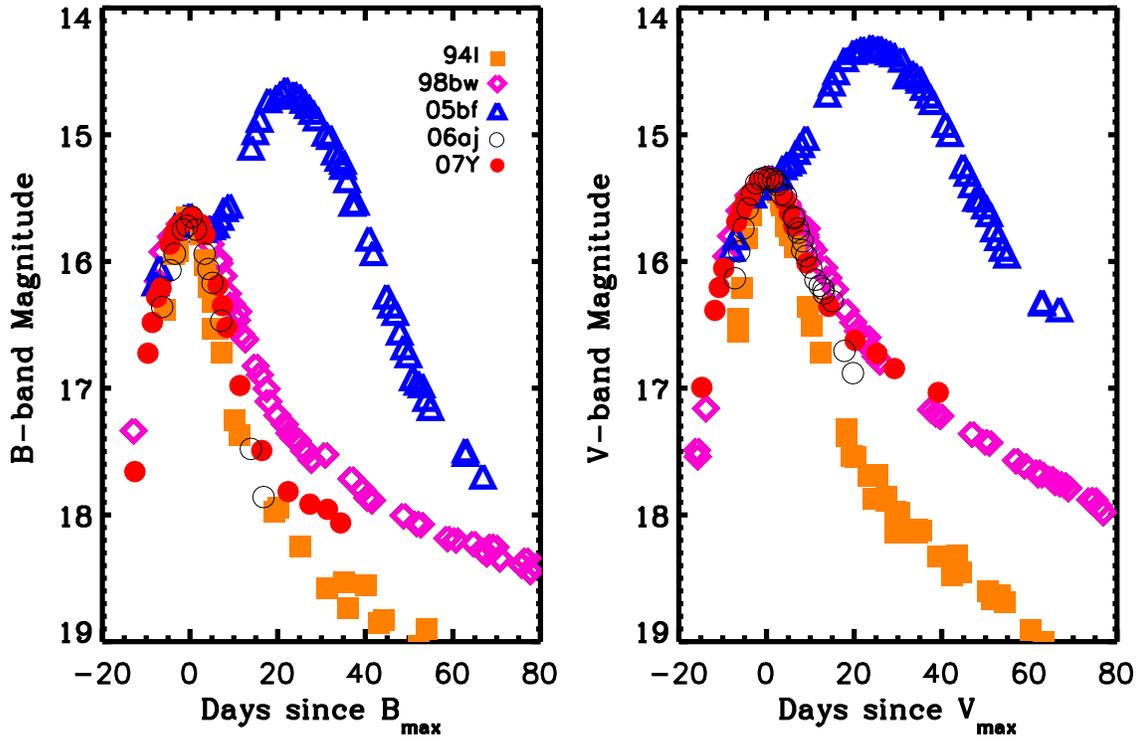}
\caption[fig6.eps]
{Comparison of the $B$- and $V$-band light curves of SN~2007Y to those of SNe~1994I, 1998bw,  2005bf and 2006aj. 
The light curves of the comparison SNe~Ib/c have been shifted to match the time and value of L$_{\rm max}$ for SN~2007Y. 
Light curves of SNe~1998bw and 2006aj have been corrected for time dilation. 
Photometry of SN~2006aj has had the host galaxy contribution removed.
SN~2007Y rises similar to SN~1998bw, however,
after maximum it evolves significantly faster. 
After $\sim$2 weeks past maximum the rate of decline
in SN~2007Y weakened causing the $V$ band to match again  
SN~1998bw.
 The first peak in SN~2005bf has a similar rise compared to SN~2007Y, but then
 evolves to another brighter secondary peak. \label{fig:photcompare}}
\end{figure}

\clearpage
\begin{figure}
\figurenum{7}
\epsscale{1.0}
\plotone{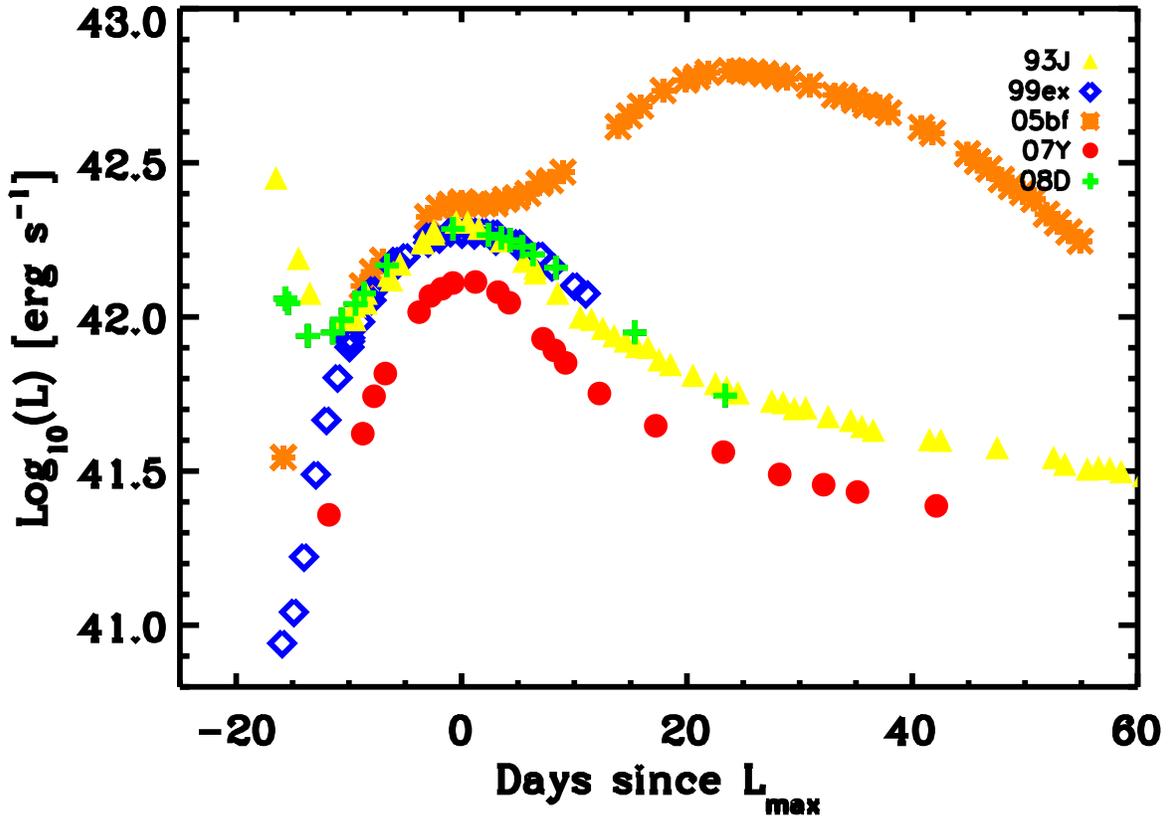}
\caption[fig7.eps]
{Comparison of the bolometric (UVOIR) light curves of SN~2007Y (red filled circles) with 
 the SNe~Ib 1999ex (blue diamonds), 2005bf (orange stars),
 2008D  (green crosses), and the SN~IIb 1993J (yellow triangles).  
A reddening of E($B$-$V$) $=$ 0.112 mag and a distance of 19.31 Mpc have been 
adopted to place the UVOIR light curve of SN~2007Y on an absolute flux scale.
Note that the last four epochs of SN~2007Y do not include any contribution from the ultraviolet. \label{fig:bolometric}}
\end{figure}

\clearpage
\begin{figure}
\figurenum{8}
\epsscale{1.0}
\plotone{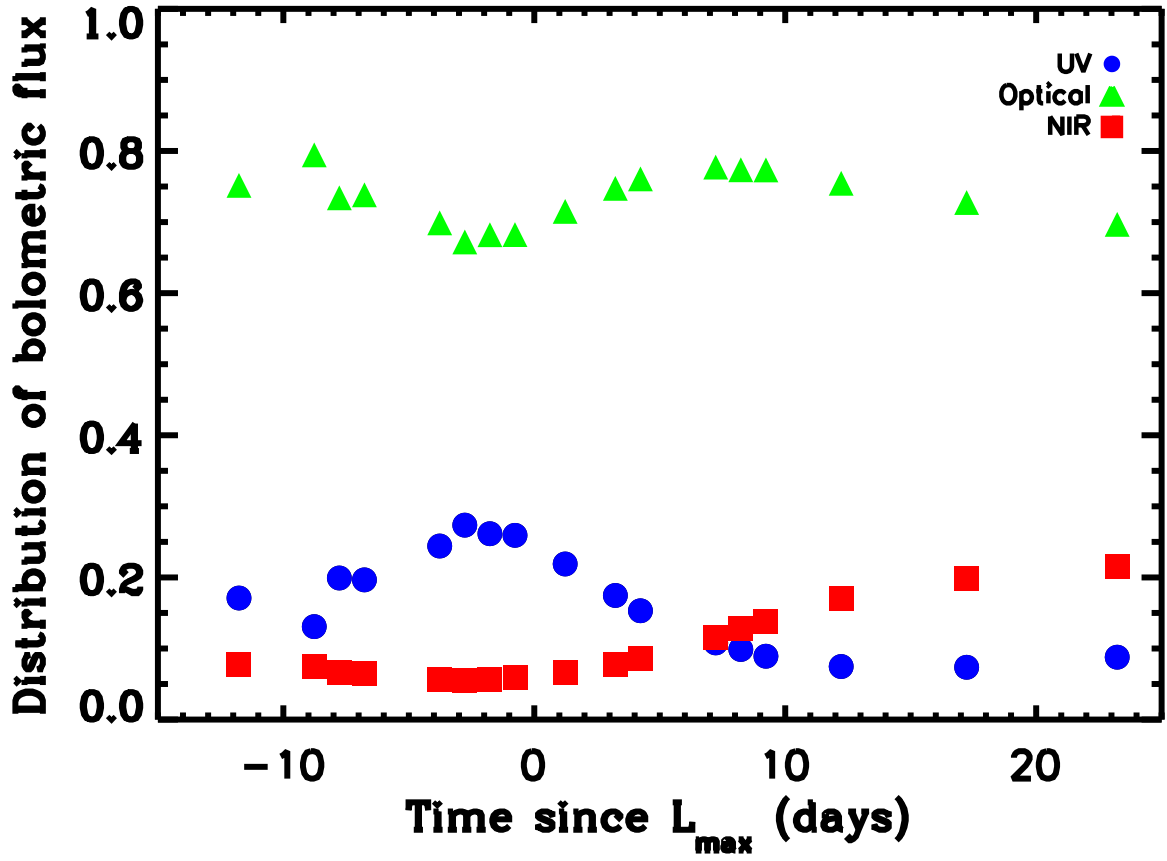}
\caption[fig8.eps]
{Distribution of flux as a function of time in three wavelength
regimes. Filled blue circles represent the flux blue-wards of the atmospheric
cutoff (3150 \AA), filled green triangles the optical flux, and
filled red squares the near-infrared ($\sim$0.95 to 2.5 $\mu$) contribution.
\label{fig:sed_compare}}
\end{figure}

\clearpage
\begin{figure}
\figurenum{9}
\epsscale{1.0}
\plotone{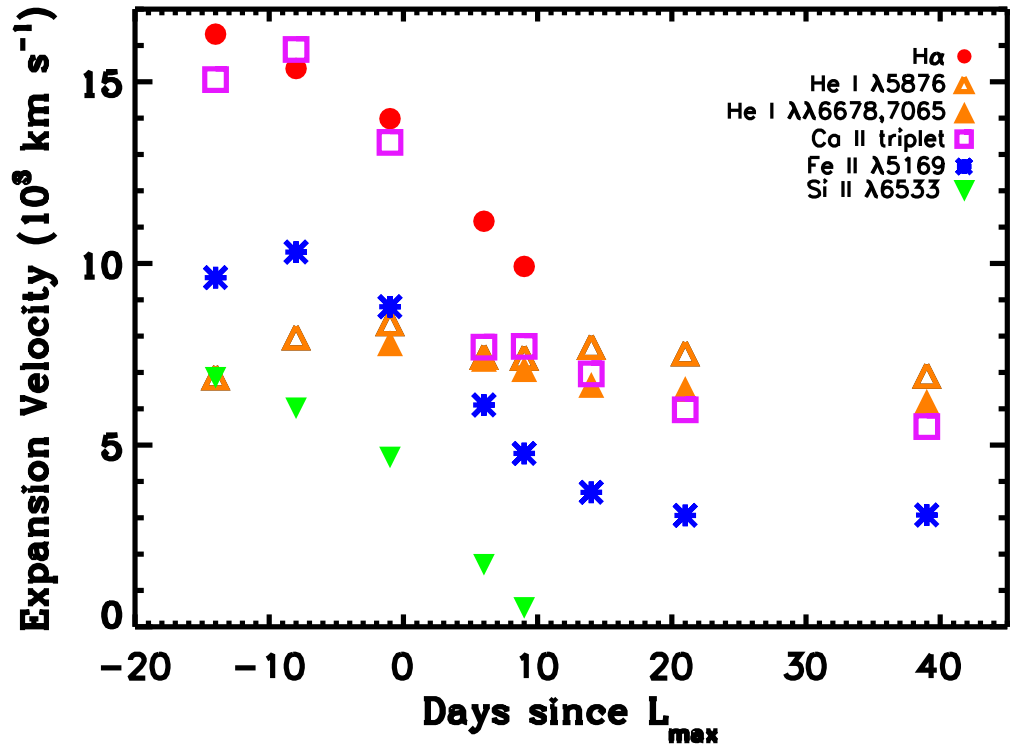}
\figcaption[fig9.eps]
{Velocity at maximum absorption of prominent ions in the spectra of SN~2007Y.  \label{fig:velocities}}
\end{figure}

\clearpage
\begin{figure}
\figurenum{10}
\epsscale{1.0}
\plotone{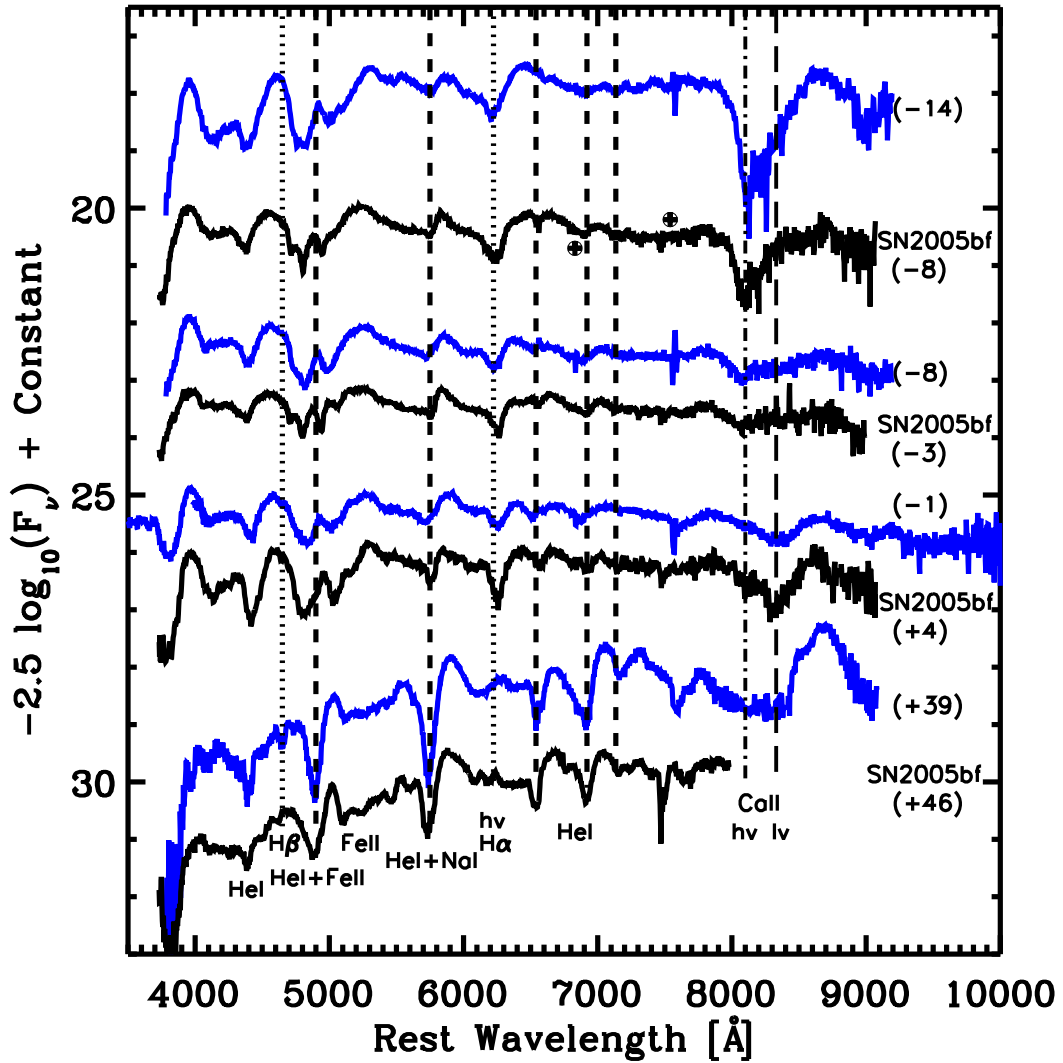}
\figcaption[fig10.eps]
{Comparison of similar epoch spectra of SN~2007Y (blue) and SN~2005bf
(black) \citep{folatelli06}.
Labels in parentheses to the right of each spectrum indicated days from L$_{\rm max}$.
In the case of SN~2005bf the time is with respect to its first maximum.
The spectra are remarkably similar and have nearly an identical evolution. 
Ions responsible for some of the main features are indicated.
Normal dash vertical line corresponds to \ion{He}{1} $\lambda\lambda$4472, 5876, 6678, 
7065, 7181; dotted lines are H$\alpha$ and H$\beta$, and high- and low- velocity components of 
\ion{Ca}{2} are represented by dot-dashed and long dashed lines, respectively. 
Telluric features are indicated with an Earth symbol.
Note that the first spectrum of SN~2007Y has been smoothed by 
averaging over 3 pixels.\label{fig:spectracompare}}
\end{figure}

\clearpage
\begin{figure}
\figurenum{11}
\epsscale{1.0}
\plotone{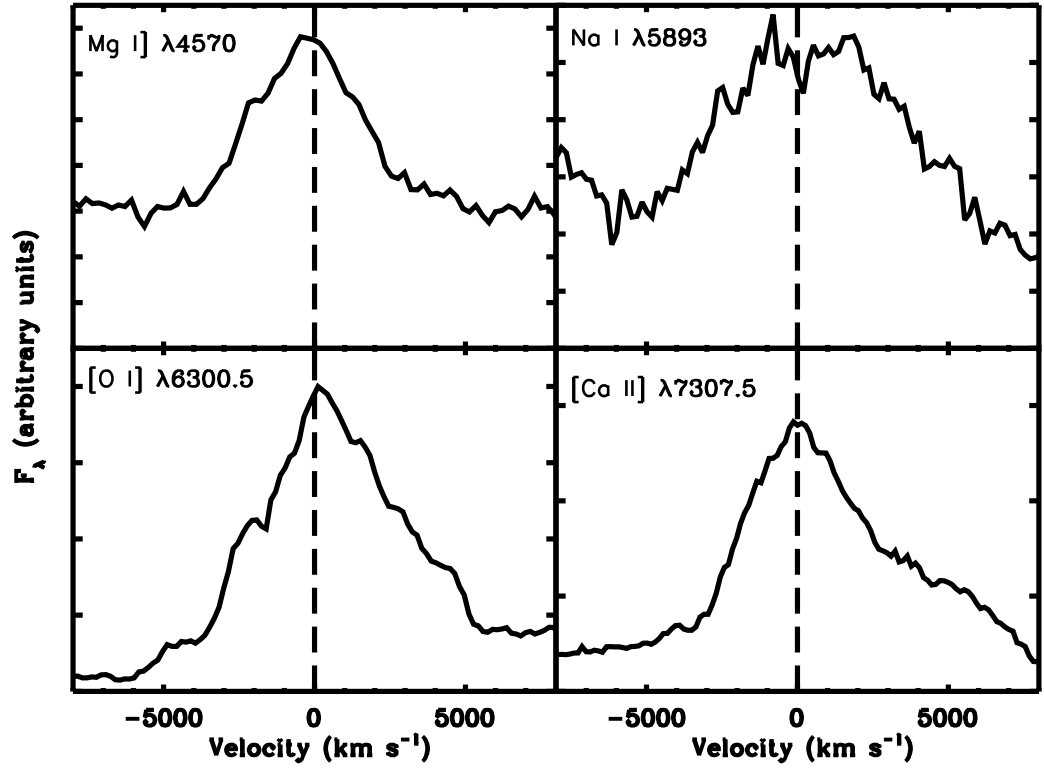}
\caption[fig11.eps]
{Prominent emission lines, in velocity space, from the  day $+270$ nebular spectrum. 
The rest wavelength of each line is marked by a vertical dashed line. 
No major asymmetries or blue-shifts are present. \label{fig:lineprofiles}}
\end{figure}

\clearpage
\begin{figure}
\figurenum{12}
\epsscale{1.0}
\plotone{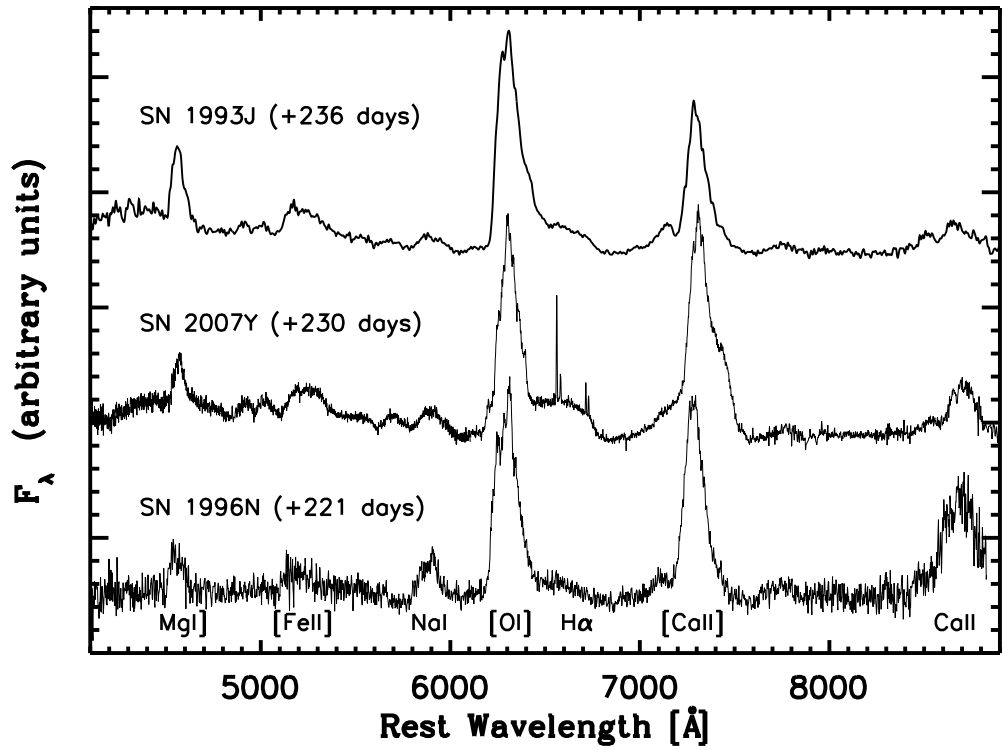}
\caption[fig12.eps]
{A comparison of similar phase nebular spectra of SN~1993J \citep{barbon94}, 
SN~1996N \citep{sollerman98}, and SN~2007Y. The wavelength scale has been 
corrected for the redshift to the host galaxy of each supernovae. Prominent
 emission features are identified.\label{fig:LTspectracompare}}
\end{figure}

\clearpage
\begin{figure}
\figurenum{13}
\epsscale{1.0}
\plotone{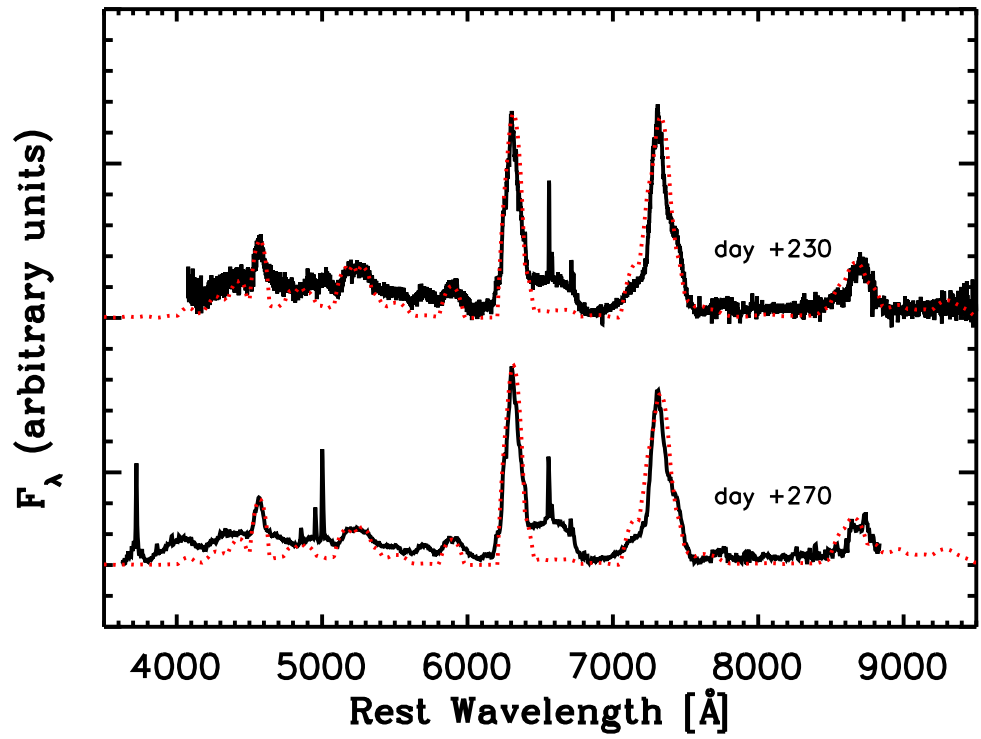}
\caption[fig13.eps]
{Nebular spectra obtained on day $+$230 and $+$270 compared to 
synthetic spectra (dotted line).\label{fig:synthesis}}
\end{figure}

\clearpage
\begin{figure}
\figurenum{14}
\epsscale{1.0}
\plotone{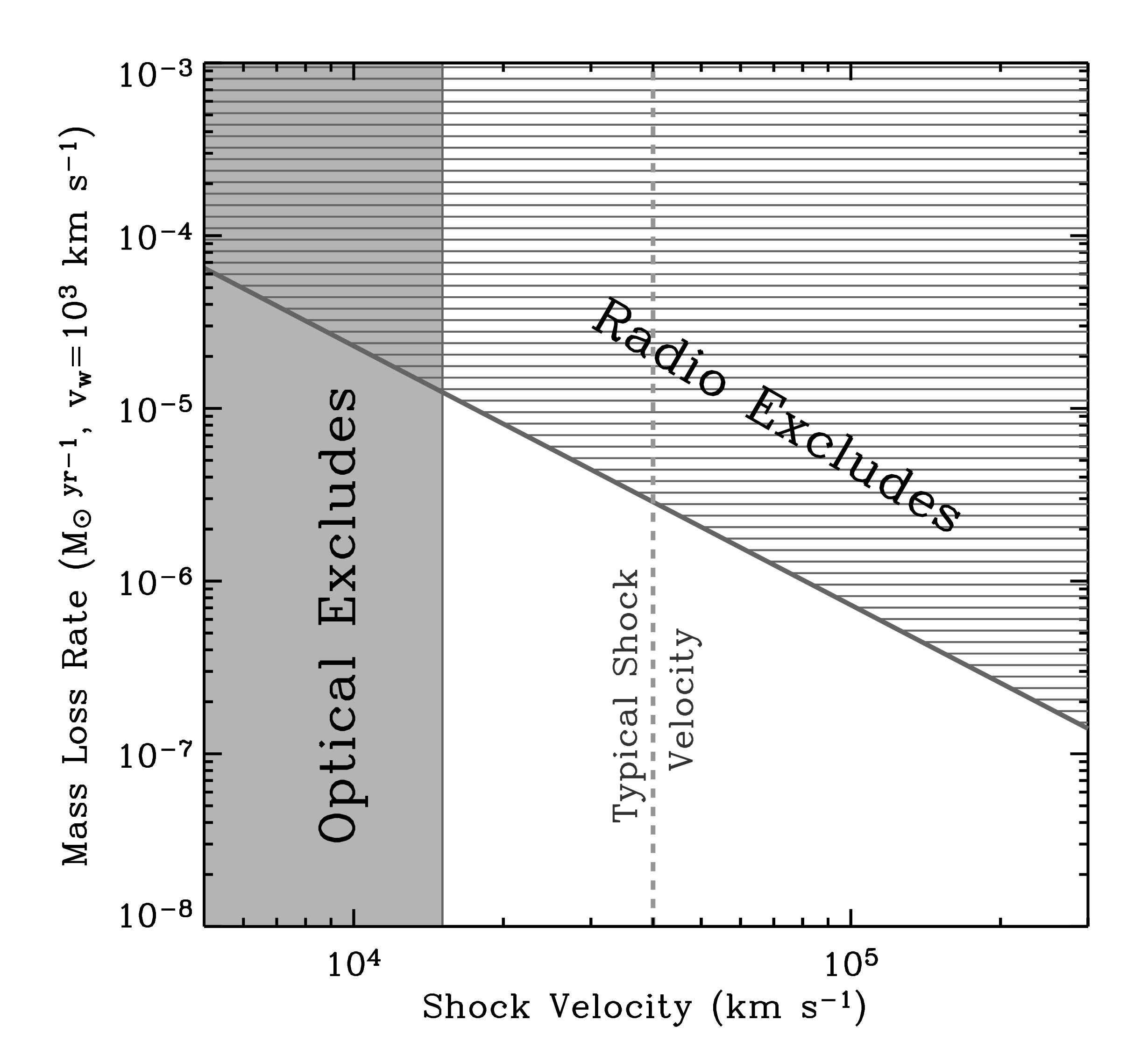}
\caption[fig14.eps]
{The two-dimensional v$_s-\dot{M}$ parameter space.
Combining constraints on the photospheric velocity derived from the 
early phase spectra and
the upper limits from radio observations we find  
a mass-loss rate of the progenitor star to be 
 $\dot{\rm M}\lesssim10^{-6}$~M$_{\odot}$~yr$^{-1}$.\label{fig:vd}}
\end{figure}

\clearpage
\begin{figure}
\figurenum{15}
\epsscale{1.0}
\plotone{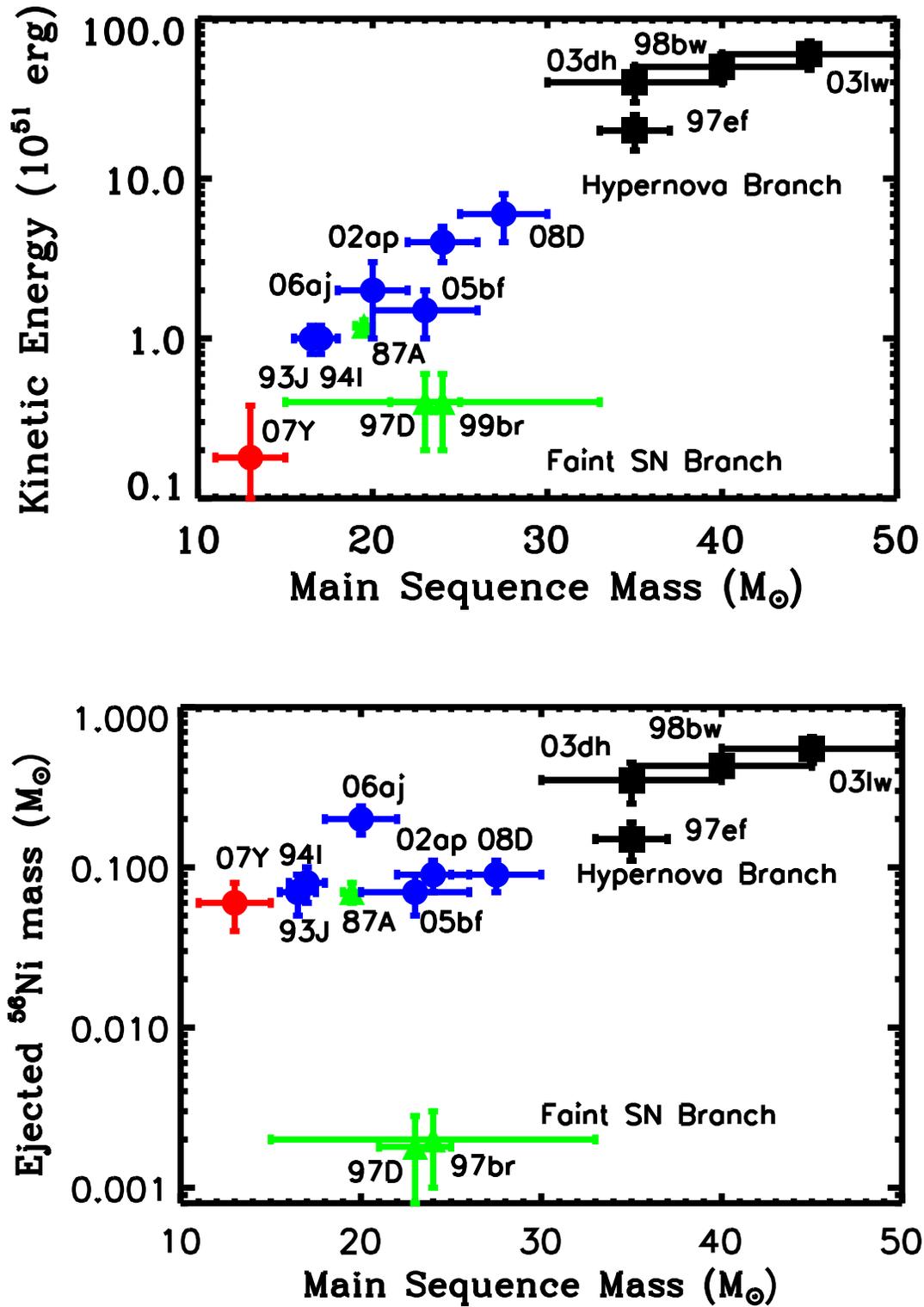}
\caption[fig15.eps]
{Comparison of physical parameters for a set of core collapse supernovae. 
{\em Upper} panel is the kinetic energy of the explosion and {\em lower} panel 
is the ejected $^{56}$Ni-mass vs. the estimated ZAMS mass of the 
progenitor stars.
Blue filled circles are normal Type~Ib/c SNe, black filled squares are Type~Ic hypernovae,
and green filled traingles are Type~II SNe. 
Parameters of the comparison supernovae  are taken from 
the literature; SN~1987A: standard values; SN~1993J: standard values; 
SN~1994I: \citet{sauer06}; 
SN~1997D: \citet{turatto98}; 
SN~1997ef: \citet{mazzali04}; 
SN~1998bw: \citet{mazzali01}; 
SN~1999br: \citet{zampieri03}; 
SN~2002ap \citet{mazzali07b};  
SN~2003dh: \citet{mazzali03}; 
SN~2003lw: \citet{mazzali06a};
SN~2005bf: \citet{tominaga05};
 SN~2006aj: \citet{mazzali06b}; 
 SN~2008D: \citet{mazzali08}. 
\label{fig:parameters}}
\end{figure}

\end{document}